\documentclass[twocolumn,showpacs,aps,floatfix,prd,nofootinbib]{revtex4}
\usepackage{graphicx}
\usepackage{dcolumn}
\usepackage{amsmath}
\usepackage{epsfig}
\usepackage{colordvi}
\usepackage{color}
\usepackage{hhline}
\usepackage{psfrag}

\RequirePackage{xspace}

\newcommand{\BABARPubYear}    {05}
\newcommand{\BABARPubNumber}  {050}
\newcommand{\SLACPubNumber} {11587}
 
\usepackage{relsize}
\def\babar{\mbox{\slshape B\kern-0.1em{\smaller A}\kern-0.1em
    B\kern-0.1em{\smaller A\kern-0.2em R}}}
     
\mathchardef\Upsilon="7107
\def\Y#1S{\ensuremath{\Upsilon{(#1S)}}\xspace}

\def\pep2{PEP-II}

\long\def\inst#1{\par\nobreak\kern 4pt\nobreak
  {\it #1}\par\vskip 10pt plus 3pt minus 3pt}
  
\begin{document}

\begin{flushleft}
SLAC-PUB-\SLACPubNumber \\
\babar-PUB-\BABARPubYear/\BABARPubNumber \\
\end{flushleft}                                                                 

\title{\large \bf
\boldmath
A study of $e^+e^-\to p\bar{p}$ using initial state radiation
with \babar\
}

\author{B.~Aubert}
\author{R.~Barate}
\author{D.~Boutigny}
\author{F.~Couderc}
\author{Y.~Karyotakis}
\author{J.~P.~Lees}
\author{V.~Poireau}
\author{V.~Tisserand}
\author{A.~Zghiche}
\affiliation{Laboratoire de Physique des Particules, F-74941 Annecy-le-Vieux, France }
\author{E.~Grauges}
\affiliation{IFAE, Universitat Autonoma de Barcelona, E-08193 Bellaterra, Barcelona, Spain }
% ...
\collaboration{The \babar\ Collaboration}
\noaffiliation

\date{\today}

\begin{abstract}

The $e^+e^-\to p\bar{p}$ cross-section is determined over a  range of
$p\bar{p}$ masses, from threshold to 4.5~GeV/$c^2$, by studying the
$e^+e^-\to p\bar{p}\gamma$ process.  The data set corresponds to an
integrated luminosity of  232~fb$^{-1}$, collected with
the \babar\ detector at the PEP-II storage ring,
at an $e^+e^-$ center-of-mass
energy of 10.6~GeV.    The mass dependence of
the ratio of electric and magnetic form factors, $|G_E/G_M|$, is
measured for $p\bar{p}$ masses below 3~GeV/$c^2$; its value
is found to be significantly larger than 1 for masses up to
2.2~GeV/$c^2$. We also measure $J/\psi \to p\bar{p}$ and
$\psi(2S) \to p\bar{p}$ branching fractions and set an upper
limit on $Y(4260)\to p\bar{p}$ production and decay.
\end{abstract}


\maketitle

\setcounter{footnote}{0}

\section{ \boldmath Introduction}
\label{intro}

\begin{figure}
\includegraphics[width=.3\textwidth]{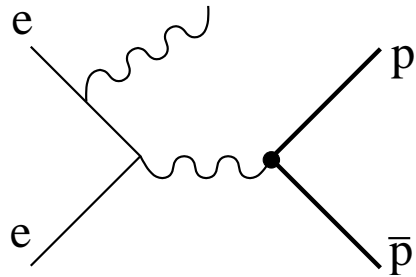}
\caption{The diagram for the $e^+e^-\to p\bar{p}\gamma$ process.
\label{diag}}
\end{figure}

The  $e^+e^-\to p\bar{p}$ 
cross-section and the proton form factor
can be measured over a range of center-of-mass energies
by studying the initial state radiation (ISR) process
$e^+e^-\to p\bar{p}\gamma$ (Fig.~\ref{diag}).  
The emission of a photon in the initial state gives rise to the 
possibility of measuring the cross-section of the 
nonradiative process $e^+e^-\to p\bar{p}$  over a range of effective 
center-of-mass energies, from the threshold
$m=2m_p=1.88$~GeV/$c^2$ to the 
full $e^+e^-$ center-of-mass energy ($\sqrt{s}$).
The Born cross-section for this process,
integrated over the nucleon momenta, is given by
\begin{equation}
\frac{{d}^2\sigma_{e^+e^-\to p\bar{p}\gamma}(m)}
{{d}m\,{d}\cos{\theta_\gamma^\ast}} =
\frac{2m}{s}\, W(s,x,\theta_\gamma^\ast)\,\sigma_{p\bar{p}}(m),
\label{eq1}
\end{equation}
where $m$ is the $p\bar{p}$ invariant mass,
$x\equiv{2E_{\gamma}^\ast}/\sqrt{s}=1-{m^2}/{s}$, 
and $E_{\gamma}^\ast$
and $\theta_\gamma^\ast$
are the ISR photon energy and polar angle, respectively,
in the $e^+e^-$ center-of-mass frame\footnote{Throughout this paper,
the asterisk denotes  quantities in the $e^+e^-$ center-of-mass frame.
All other variables except $\theta_p$ and $\theta_K$ are defined in the
laboratory frame.}.
The function    $W(s,x,\theta_{\gamma}^\ast)$~\cite{BM}, 
\begin{equation}
W(s,x,\theta_{\gamma}^\ast)=
\frac{\alpha}{\pi x}\left(\frac{2-2x+x^2}{\sin^2\theta_{\gamma}^\ast}-
\frac{x^2}{2}\right),
\label{eq2}
\end{equation}
is the probability of ISR photon emission for 
$\theta_{\gamma}^\ast\gg m_e/\sqrt{s}$, where
$\alpha$ is the fine-structure constant and $m_e$ is the electron mass.
The cross-section for the $e^+e^-\to p\bar{p}$ process is given by
\begin{equation}
\sigma_{p\bar{p}}(m) = \frac{4\pi\alpha^{2}\beta C}{3m^2}
\left [|G_M(m)|^{2} + \frac{2m_p^2}{m^2}|G_E(m)|^{2}\right],
\label{eq4}
\end{equation}
with $\beta = \sqrt{1-4m_p^2/m^2}$, 
$C = y/(1-e^{-y})$, 
and $ y = {\pi\alpha m_p}/({\beta m})$ 
is the Coulomb correction factor~\cite{Coulomb}, which 
makes the cross-section nonzero at threshold.
The cross-section depends on the magnetic form factor ($G_M$) and
the electric form factor ($G_E$); at threshold, $|G_{E}| = |G_{M}|$.
The modulus of the ratio of electric and magnetic form factors
can be determined from  the distribution of $\theta_p$,
the angle between the proton momentum in the $p\bar{p}$
rest frame and the momentum of the $p\bar{p}$ system in the $e^+e^-$ 
center-of-mass frame.
This distribution can
be expressed as a sum of terms proportional to
$|G_M|^2$ and $|G_E|^2$. 
The full differential cross-section for
$e^+e^-\to p\bar{p}\gamma$  can be found, for example,
in Ref.~\cite{kuhn_pp}. The $\theta_p$ dependencies of
the $G_E$ and $G_M$ terms are reminiscent of
the  $\sin^{2}\theta_p$ and
$1+\cos^{2}\theta_p$  angular distributions for electric and
magnetic form factors in the  $e^+e^-\to p\bar{p}$ process.

   Measurements of the $e^+e^-\to p\bar{p}$ cross-section
have been performed in $e^+e^-$
experiments ~\cite{DM1,DM2,ADONE73,FENICE,BES,CLEO} 
with (20--30)\% precision.
The cross-section and proton form factor were deduced
assuming  $|G_E|=|G_M|$, and the measured proton
angular distributions \cite{DM2,FENICE} did not contradict
this assumption.
More precise measurements of the proton form factor 
have been performed in
$p\bar{p}\to e^+e^-$ experiments~\cite{LEAR,E760,E835}.
In the PS170 experiment~\cite{LEAR} at LEAR,
the proton form factor was measured from threshold
($p\bar{p}$ annihilation at rest) up to a  mass of 2.05~GeV/$c^2$. The
ratio  $|G_E/G_M|$ was measured using the angular dependence of the
cross-section and was found to be compatible with unity.
The LEAR data show a strong dependence of the form factor
on  $p\bar{p}$ mass near threshold, and very little dependence
in the range 1.95--2.05~GeV/$c^2$. Analyses from
Fermilab experiments E760~\cite{E760}
and E835~\cite{E835} show a strong decrease in the form factor
at higher masses, in agreement with perturbative QCD, which predicts
a $\alpha_s^2(m^2)/m^4$ dependence.

This work is an independent measurement
by the \babar\ Collaboration of
the $e^+e^- \to p\bar{p}$ cross-section $\sigma_{p\bar{p}}(m)$,
for $p\bar{p}$ masses up to 4.5~GeV/$c^2$, based on
the ISR process in $e^+e^-$ annihilation
at a fixed center-of-mass energy near 10.6~GeV.
This study  significantly improves the measurement
of $\sigma_{p\bar{p}}(m)$ in the $p\bar{p}$ mass range up to 3~GeV/$c^2$.
In contrast to previous $e^+e^-$ and $p\bar{p}$ experiments,
our measurement does not use the assumption that $|G_E|=|G_M|$.
The ISR approach provides Á full $\theta_p$ coverage and hence
high sensitivity to $|G_E/G_M|$. In this work,
the mass dependence of the form-factor ratio $|G_E/G_M|$ is
measured for $p\bar{p}$ masses below 3~GeV/$c^2$.
We also study $J/\psi$ and $\psi(2S)$ production in
$e^+e^-\to p\bar{p}\gamma$, and measure the products
$\Gamma(\psi\to e^+e^-){\cal B}(\psi\to p\bar{p})$.
A search for production of the $Y(4260)$ resonance, recently observed
by \babar\ in the ISR process
$e^+e^-\to Y(4260)\gamma \to J/\psi \pi^+\pi^- \gamma$~\cite{yexp},
is performed.

\section{ \boldmath The \babar\ detector and data samples}
\label{detector}
We analyse a data sample corresponding to
232~fb$^{-1}$ recorded with
the  \babar\ detector~\cite{ref:babar-nim} at the \pep2\ 
asymmetric-energy storage ring. At \pep2, 9-GeV electrons collide with 
3.1-GeV positrons at a center-of-mass energy of 10.6~GeV 
(the $\Upsilon$(4S) resonance).

Charged-particle tracking is
provided by a five-layer silicon vertex tracker (SVT) and
a 40-layer drift chamber (DCH), operating in a 1.5-T axial
magnetic field. The transverse momentum resolution
is 0.47\% at 1~GeV/$c$. Energies of photons and electrons
are measured with a CsI(Tl) electromagnetic calorimeter
(EMC) with a resolution of 3\% at 1~GeV. Charged-particle
identification is provided by specific ionization (${d}E/{d}x$) 
measurements in the SVT and DCH, and by an internally reflecting 
ring-imaging Cherenkov detector (DIRC). Muons are identified
in the solenoid's instrumented flux return,
which consists of iron plates interleaved with resistive
plate chambers.

Signal and background ISR processes are simulated with
Monte Carlo (MC) event generators based on 
Ref.~\cite{EVA}, with the  differential cross-section for 
$e^+e^-\to p\bar{p} \gamma$ taken from Ref.~\cite{kuhn_pp}.
   Because the polar-angle distribution of the ISR photon is peaked
near $0^\circ$ and $180^\circ$, the MC events are generated with
a restriction on the photon polar angle: 
$20^\circ<\theta_{\gamma}^\ast<160^\circ$, where 
$\theta_{\gamma}^\ast$ is measured in the $e^+e^-$ center-of-mass frame.
The extra soft-photon radiation from the initial state is generated 
with the structure function method~\cite{strfun}. To restrict the 
maximum energy of the extra photons, the 
invariant mass of the hadron system combined with the ISR photon
is required to be at least 8~GeV/$c^2$.    For background
$e^+e^- \to \mu^+\mu^-\gamma$, $\pi^+\pi^- \gamma$, and $K^+K^-\gamma$
processes, final state Bremsstrahlung is generated using the  PHOTOS
package~\cite{PHOTOS}.
Background from $e^+e^- \to q\bar{q}$
   is simulated with the JETSET~\cite{JETSET} event generator. 
   The response of
   the \babar\ detector is simulated using the GEANT4~\cite{GEANT4} program.
The simulation takes into account the variation of the detector and
accelerator conditions, and   beam-induced
background  photons and charged particles overlapping  events of interest.

\section{ \boldmath Event selection}
\label{selection}
The preselection of $e^+e^- \to p\bar{p}\gamma$ candidates
requires that all the final-state particles are detected inside a 
fiducial volume. Since a significant fraction of the events
contain beam-generated spurious tracks and photon candidates, we
select events with at least two tracks with opposite charge and
at least one photon candidate with $E^\ast_\gamma>3$~GeV.
The polar angle of the photon is required to be in the well-understood
region of the calorimeter:
$21.5^\circ <\theta_\gamma < 137.5^\circ$.
The charged tracks must originate from the interaction point, have
transverse momentum greater than 0.1~GeV/$c$, and be in the angular
region between $25.8^\circ$ and $137.5^\circ$, so that
particle identification (PID) may be performed using the DIRC detector.
To suppress background from radiative Bhabha events, 
events in which each of the two highest momentum tracks has a  ratio 
of calorimetric energy deposition to momentum in the range 0.9 to 1.1 
are rejected.

For events passing the preliminary selection, a kinematic fit is
 performed to the $e^+e^- \to C^+C^- \gamma$ hypothesis with requirements of
  total energy and momentum conservation. Here $C$ can be
$e$, $\mu$, $\pi$, $K$ or $p$, and $\gamma$ is the photon candidate with the
highest energy in the $e^+e^-$ center-of-mass frame. 
For events with more than two charged tracks,
the fit uses the two oppositely charged tracks 
that pass closest
to the interaction point. 
The Monte Carlo simulation does not accurately reproduce the shape
of the photon energy resolution function. This leads to a
difference in the distributions of the $\chi^2$ of the kinematic
fit for data and for MC simulated events. To reduce this difference, only
the measured direction of the ISR photon is used in the fit;
its energy is treated as a free fit parameter.
For each of the  five charged-particle mass hypotheses,
the corrected angles and  energies of the particles and the $\chi^2$ of
the kinematic fit are calculated. 

The selection of $e^+e^-\to p\bar{p}\gamma$ events relies upon
both particle identification and event kinematics. The expected 
number of events from the background processes
       $e^+e^-\to \pi^+ \pi^- \gamma$, $\mu^+\mu^- \gamma$, and
$K^+K^- \gamma$ significantly  exceeds the number of signal events 
(by two to three orders of magnitude).  To suppress these backgrounds, 
both charged particles must be identified as protons according to the
specific ionization (dE/dx) 
measured in the SVT and DCH, and the Cherenkov angle measured
in the DIRC. 
These particle identification 
requirements lead to a loss of approximately 30\% of the signal events,
while suppressing backgrounds by   factors of $15\times 10^3$, 
$500\times 10^3$, and $2\times 10^3$ 
for pion, muon, and kaon events, respectively.

\begin{figure}
\includegraphics[width=.48\linewidth]{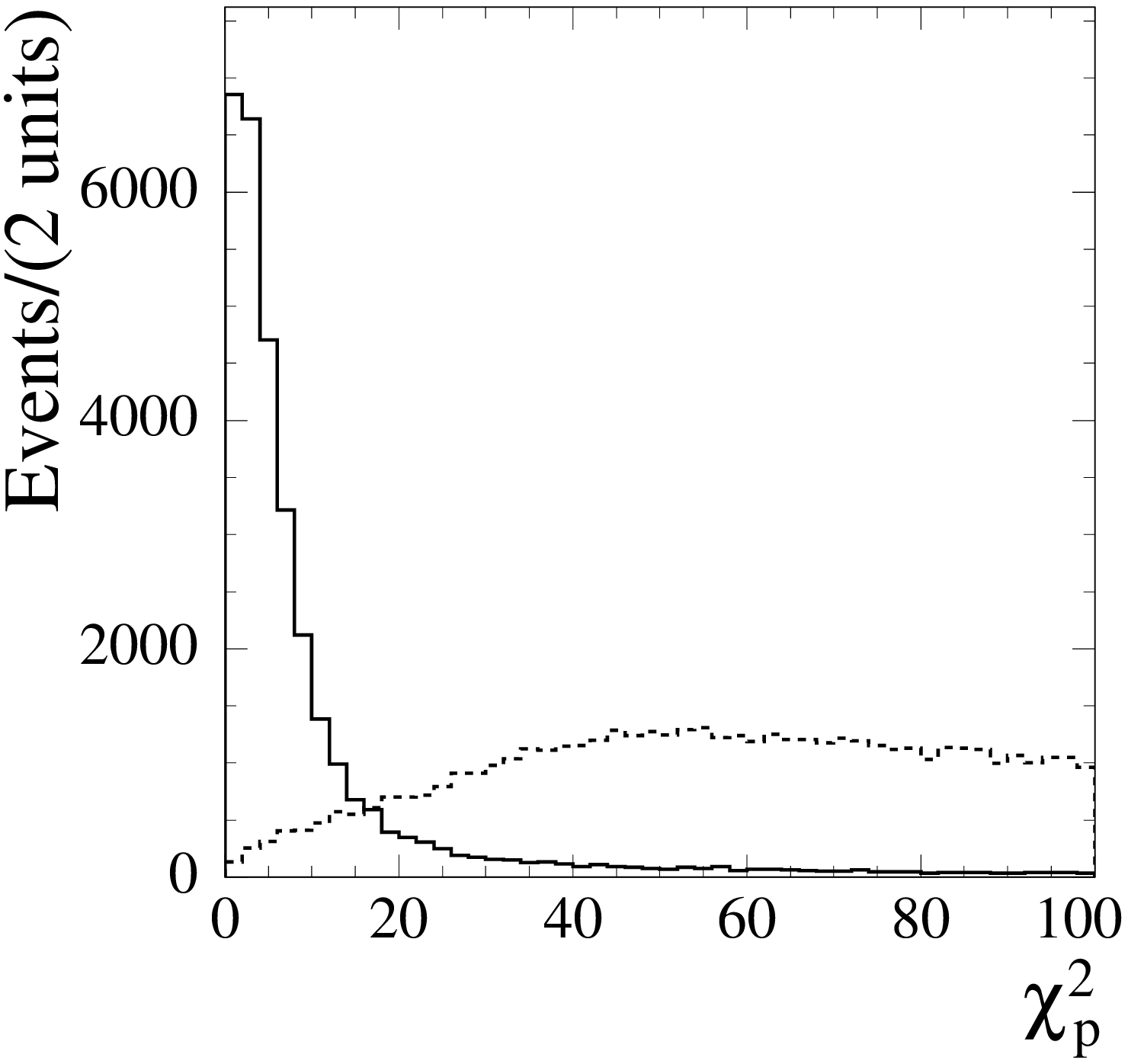}
\hfill
\includegraphics[width=.48\linewidth]{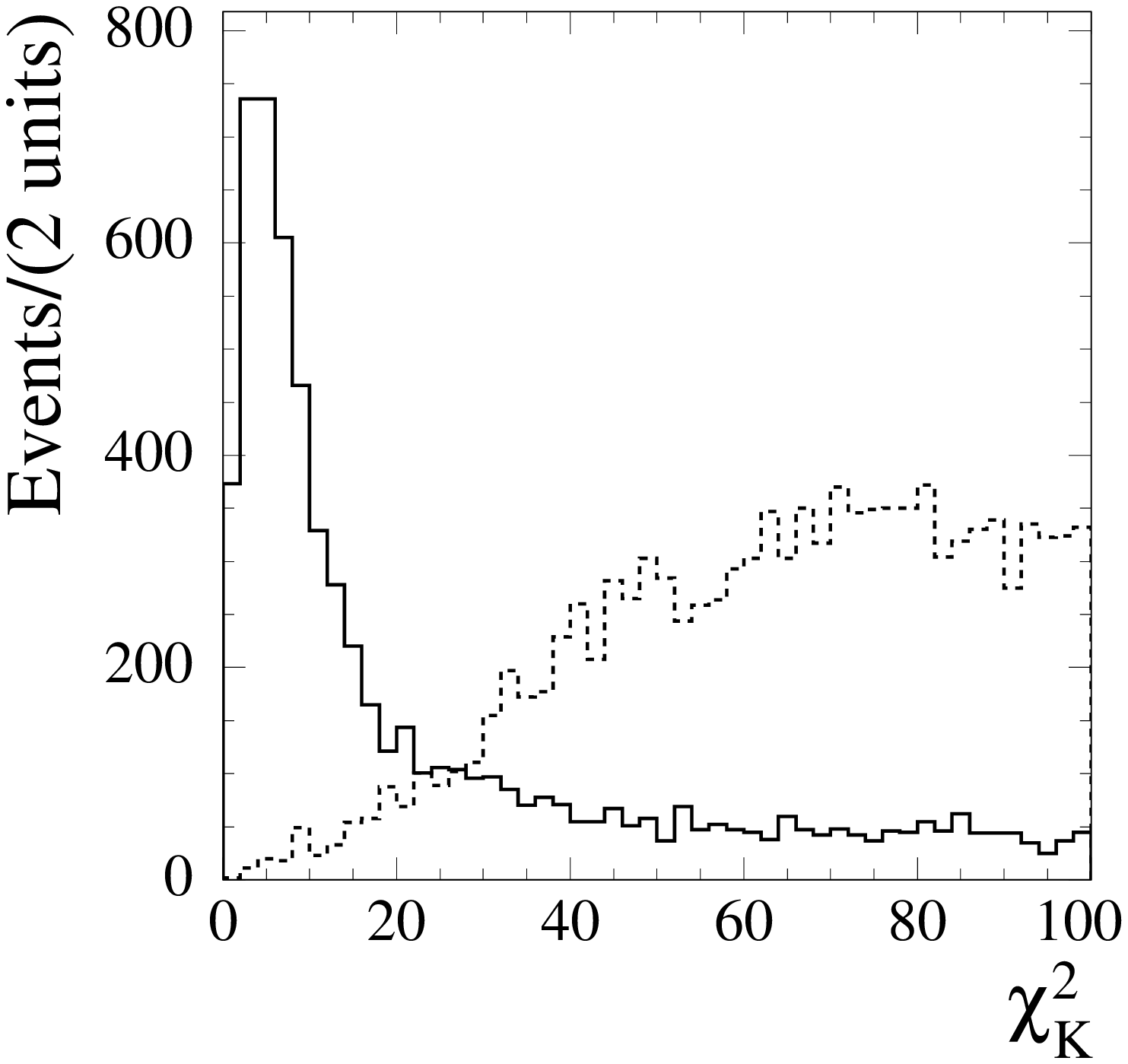}
\caption{The $\chi^2_p$ distribution (left) for MC simulated
$e^+e^-\to p \bar{p}\gamma$ (solid line) and
$e^+e^-\to K^+K^-\gamma$ (dashed line) events, and the
the $\chi^2_K$ distribution  (right) for  MC simulated
$e^+e^-\to K^+K^-\gamma$ (solid line) and
$e^+e^-\to p \bar{p}\gamma$ (dashed line) events with $\chi^2_p<30$.
\label{chi_ppg}}
\end{figure}

Background is further suppressed through requirements on the
$\chi^2$ of the kinematic fit:
$\chi^2_p<30$ and $\chi^2_K>30$, where  $\chi^2_p$ and
$\chi^2_K$ are the $\chi^2$ of the kinematic fit for the
proton and kaon mass hypotheses, respectively.
The distribution of $\chi^2_p$ for Monte Carlo simulated 
$p\bar{p}\gamma$ events is shown in Fig.~\ref{chi_ppg}~(left). 
The long tail in the distribution at high $\chi^2$ is due to 
events with extra photons emitted in the initial state. 
The dashed histogram is the $\chi^2_p$ distribution for 
$K^+K^-\gamma$ Monte Carlo simulated  events. 
Figure~\ref{chi_ppg}~(right) shows the distributions
of $\chi^2_K$ for $K^+K^-\gamma$  and $p\bar{p}\gamma$
Monte Carlo simulated events with $\chi^2_p<30$.
The $\chi^2$ requirements lead to a loss of 25\% of  signal events
but provides additional background suppression by a factor of
50 for  $e^+e^-\to \pi^+ \pi^- \gamma$ and $\mu^+\mu^- \gamma$ events,
and a factor of 30 for $e^+e^-\to K^+K^-\gamma$ events.

\begin{figure}
\psfrag{pp}{$\mbox{p}\bar{\mbox{p}}$}
\includegraphics[width=.45\textwidth]{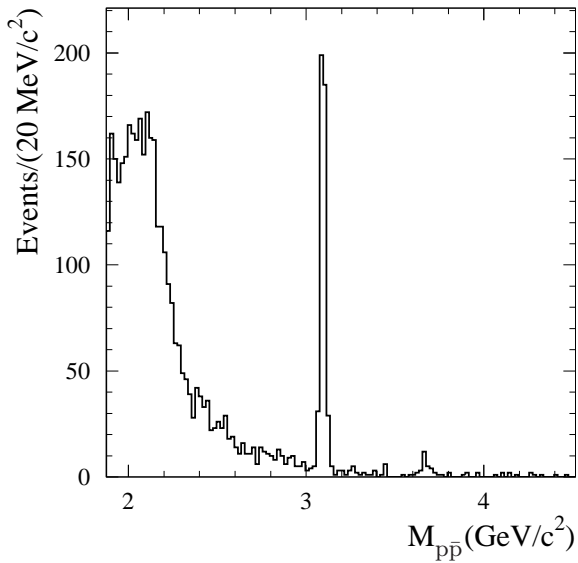}
\caption{The $p\bar{p}$ invariant mass spectrum for
$p\bar{p}\gamma$ candidates that satisfy all selection criteria.
\label{rawspec}}
\end{figure}

The  $p\bar{p}$ invariant mass distribution 
is shown in Fig.~\ref{rawspec}
for the $\approx4000$ events that satisfy all the selection criteria.
Most of the events have invariant mass below 3~GeV/$c^2$.
Clear signals from $J/\psi\to p\bar{p}$ and $\psi(2S)\to p\bar{p}$ 
decays are evident.

\section{ \boldmath Background and its subtraction}
\label{subtraction}
The possible sources of background in the sample of 
$e^+e^-\to p\bar{p}\gamma$ candidates
that pass the selection criteria described in the 
previous section include
$e^+e^-\to \pi^+\pi^-\gamma$,  $e^+e^-\to K^+K^-\gamma$,
$e^+e^-\to \mu^+\mu^-\gamma$, and $e^+e^-\to e^+e^-\gamma$
events in which the charged particles are misidentified as protons.
Backgrounds from processes with protons plus neutral particle(s) 
in the final state are also anticipated: 
$e^+e^-\to p\bar{p}\pi^0$, $p\bar{p}\eta$, $p\bar{p}\pi^0\gamma$, etc.

Of particular interest is the possible background from the process 
$e^+e^-\to p\bar{p}\gamma$ with the photon emitted from the final state.
Due to different charge parity of the amplitudes corresponding to
initial state radiation and final state  radiation (FSR), 
their interference does not
contribute to the total $e^+e^-\to p\bar{p}\gamma$ cross-section.
The contribution of the FSR amplitude is estimated to be~\cite{Chernyak}
${\rm d}\sigma/{\rm d} m\approx |F_{\rm ax}|^2{8m\alpha^{3}\beta}/({27s^2})$,
where $F_{\rm ax}$ is the axial proton form factor.
Assuming $|F_{\rm ax}|\approx |G_M|$,  the ratio of FSR to ISR cross-sections 
is determined to be about $10^{-3}$ for 
$p \bar{p}$ masses below 4.5~GeV/$c^2$, 
implying that the FSR background is sufficiently small to 
neglect.
 
\subsection{\boldmath $e^+e^-\to \pi^+\pi^-\gamma$, 
$e^+e^-\to K^+ K^-\gamma$, $e^+e^-\to \mu^+\mu^-\gamma$ and 
$e^+e^-\to e^+e^-\gamma$ backgrounds}

 To estimate the background contribution from $e^+e^-\to \pi^+\pi^-\gamma$,
data and Monte Carlo simulated events are selected with the
following requirements on PID and on the $\chi^2$ of the kinematic fits:
\begin{enumerate}
\item one proton candidate, $\chi^2_{\pi}<20$;
\item one proton candidate, $\chi^2_{p}<30$, $\chi^2_{K}>30$;
\item two proton candidates, $\chi^2_{\pi}<20$;
\item two proton candidates, $\chi^2_{p}<30$, $\chi^2_{K}>30$.
\end{enumerate}
Here $\chi^2_{\pi}$ is the $\chi^2$ of the kinematic fit for
the pion mass hypothesis.

\begin{figure}
\includegraphics[width=.48\linewidth]{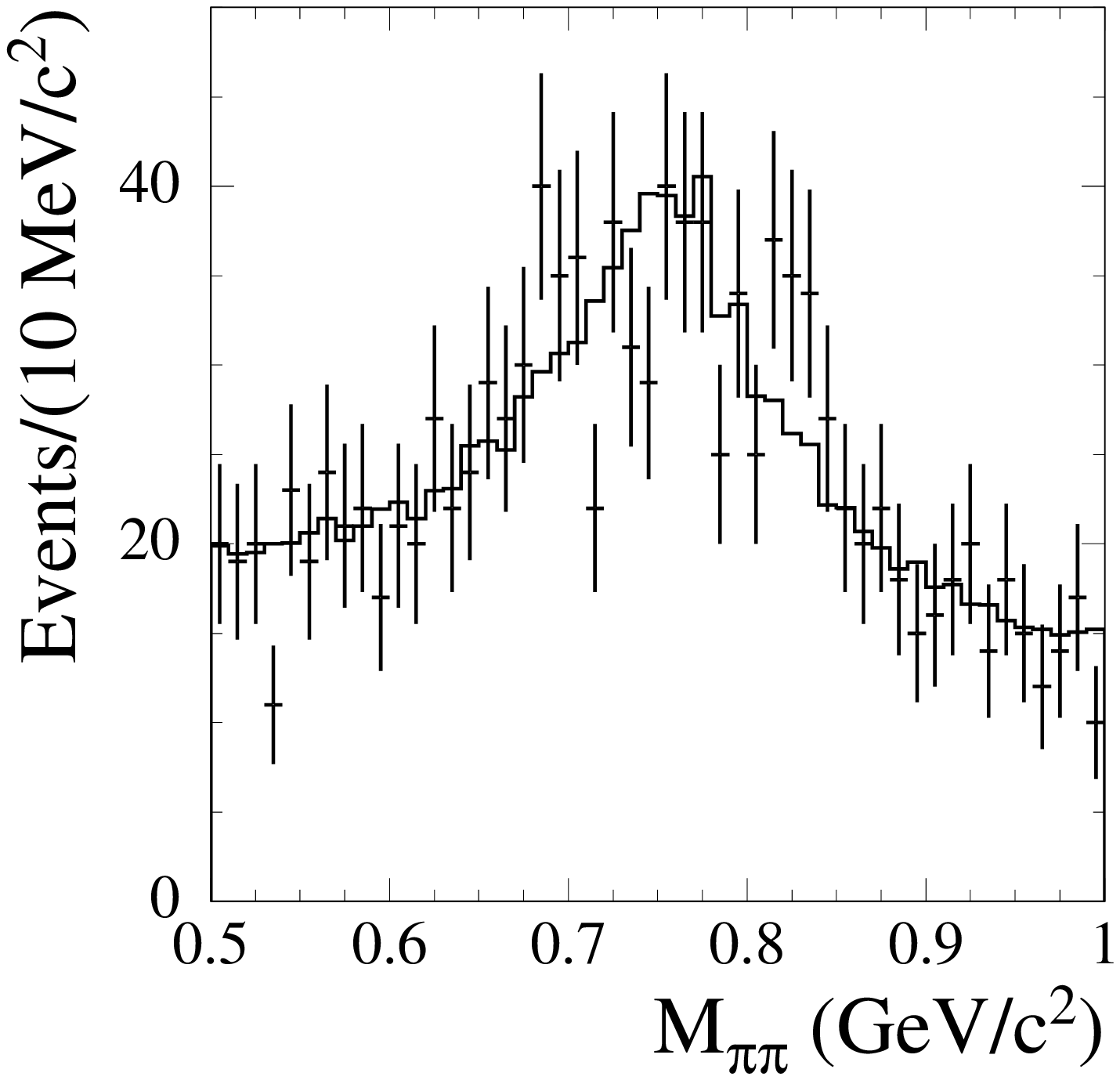}
\hfill
\includegraphics[width=.48\linewidth]{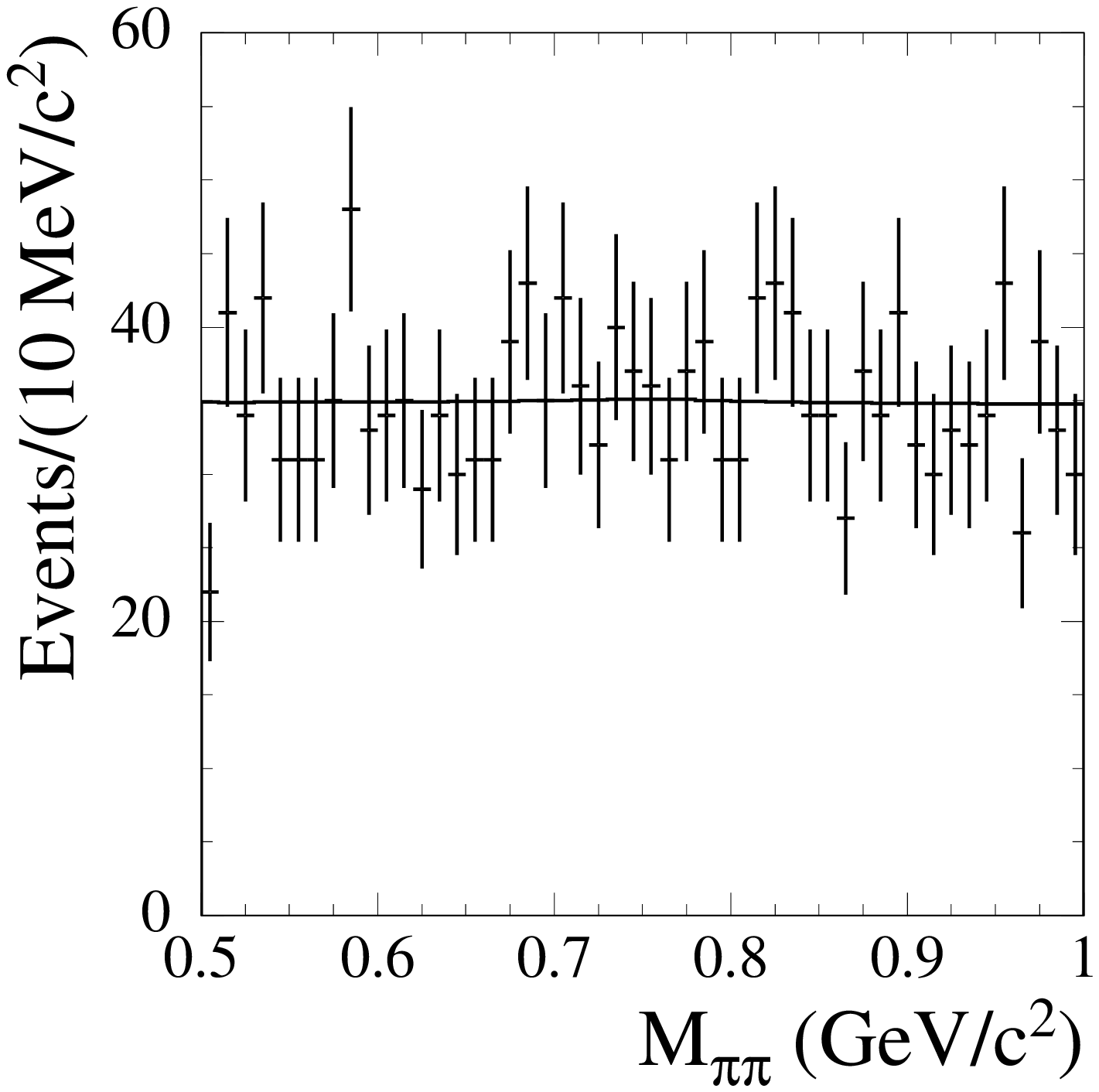}
\caption{The $M_{\pi\pi}$ spectrum for data  events with
$\chi^2_{p}<30$ and $\chi^2_{K}>30$,
and one proton candidate (left plot; selection 2 in the text) 
or two proton candidates (right plot; selection 4 in the text).
The histograms are the results of the fit described in the text.
\label{pibkg}}
\end{figure}

The fourth set of conditions corresponds to the standard selection criteria for
$p\bar{p}\gamma$ candidates.
The invariant mass $M_{\pi\pi}$ of the two charged particles
under the pion-mass hypothesis is calculated;
the $M_{\pi\pi}$ distributions for data  selected
with criteria 2 and 4 are shown in Fig.~\ref{pibkg}.
The $\rho$ resonance in the $e^+e^-\to \pi^+\pi^-\gamma$ reaction
is clearly seen in
the distribution corresponding to selection 2 (left plot in Fig.~\ref{pibkg}.
The number of $\pi\pi\gamma$ events with $0.5<M_{\pi\pi}<1$ GeV/$c^2$
passing each set of selection criteria
is determined by fitting the $M_{\pi\pi}$ distribution with a
$\pi\pi\gamma$ spectrum predicted by Monte Carlo plus a first
order polynomial to account for background from non-$\pi\pi\gamma$              
processes.
      The Monte Carlo $\pi\pi\gamma$ spectrum uses a
model of the pion form factor based on existing experimental 
data. The results of the fits for $\pi\pi\gamma$ candidates passing 
selection criteria 1, 2, 3 and 4 are listed
in Table~\ref{pibkg_tab} together with the corresponding numbers from 
the Monte Carlo  simulation.
\begin{table*}
\caption{The numbers of $\pi\pi\gamma$ events with
$0.5<M_{\pi\pi}<1$~GeV/$c^2$ passing different selection
criteria for data and Monte Carlo simulation (MC).
WMC denotes Monte Carlo simulation
with data-derived  particle identification weights.
The data numbers are obtained from the fit of the $M_{\pi\pi}$ distributions
as described in the text.
$R_{\pi\pi}$ is the ratio of the numbers of events in the previous two rows.
\label{pibkg_tab}}
\begin{ruledtabular}
\begin{tabular}{lcccccc} 
&\multicolumn{3}{c}{1 proton candidate}&\multicolumn{3}{c}{2 proton candidates}\\
&data&MC&WMC&data&MC&WMC\\
\hline
$N(\chi^2_{\pi}<20)$&$16200\pm200$&$21020\pm230$&$12300
\pm300$&$190\pm30$&$246\pm25$&$35.5\pm0.8$\\
$N(\chi^2_{p}<30,\chi^2_{K}>30)$&$460\pm120$&$590
\pm40$&$300\pm5$&--&$<5.7$&$0.90\pm0.03$\\
$R_{\pi\pi}$&$35\pm9$&$36\pm2$&$43\pm1$&--&$>43$&$39\pm2$\\
\end{tabular}
\end{ruledtabular}
\end{table*}

\begin{figure}
\psfrag{pp}{$\mbox{p}\bar{\mbox{p}}$}
\includegraphics[width=.45\textwidth]{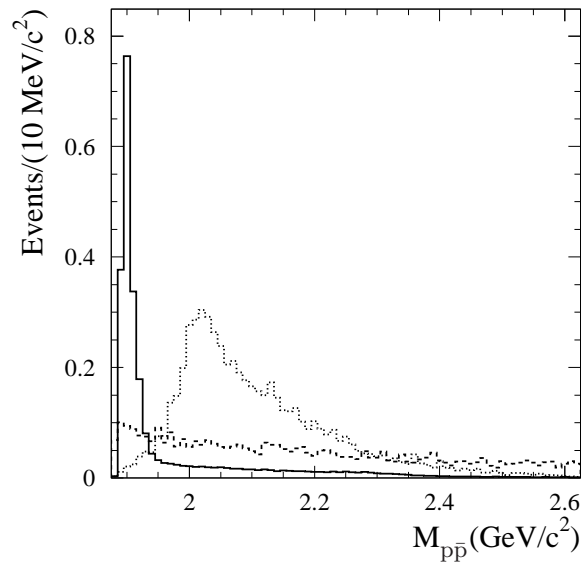}
\caption{The calculated $M_{p\bar{p}}$ spectra for 
$e^+e^-\to\pi^+\pi^-\gamma$ (dotted histogram),
$e^+e^-\to K^+ K^-\gamma$ (solid histogram), and
$e^+e^-\to \mu^+ \mu^-\gamma$ (dashed histogram)
background processes.
The spectra are
normalized to the number of events
expected to
pass the $p\bar{p}\gamma$ selection criteria
for each process:
$5.9\pm2.5$ for pions and $2.5\pm1.0$ for kaons.
For the muon channel, 
the upper limit of 11 events is used for the normalization.
\label{pipi_m}}
\end{figure}

Particle identification for the simulated $\pi\pi\gamma$ events is
accomplished using two sets of information:
fully simulated observables that are used for particle 
identification in the same manner as in the analysis of data,  
and
event weights for the simulated events based on pion 
misidentification rates derived from a control sample
of known pions in data.
The identification based on event weights does not 
take into account  possible correlations between pion misidentification 
probabilities for two  particles that overlap 
in the detector, or are in close proximity 
and therefore may underestimate the yield of wrongly identified 
$\pi\pi\gamma$ events. No events passing selection 4 are found in the 
fully simulated particle-ID sample, and so a
90\% confidence level (CL) upper limit is estimated for the standard selection
(selection 4) in Table~\ref{pibkg_tab}.
Because neither the  fully simulated nor weighted PID samples
predict  the number of
$\pi\pi\gamma$ events seen in the data passing selection 4,
an estimate is made based on the number of data
events passing selection 3:
$N_4=N_3/R_{\pi\pi}$,
where $R_{\pi\pi}$ is the ratio of the number of candidates that satisfy 
$\chi^2_\pi<20$ to the number that satisfy
$\chi^2_p<30$ and $\chi^2_K>30$ (the numbers given in the first and 
second rows in Table~\ref{pibkg_tab}.
The statistical uncertainty on the scale factor $R_{\pi\pi}$
from the simulation is about
20\%.
$R_{\pi\pi}$ estimated for events with
one and two misidentified pions are consistent with each other. 
Accordingly, the scale factor ratio $R_{\pi\pi}=35\pm9$
obtained from data is used, with an additional 30\% systematic 
uncertainty assigned.
Finally, $N_4$ is estimated as $N_4=(190\pm30)/(35\pm14)=5.4\pm2.3$.
The fit with $N_4=5.4$, shown in Fig.~\ref{pibkg}, describes
the mass distribution for selection 4  very well.
The total number of $\pi\pi\gamma$ events remaining for  the  standard
selection criteria is calculated as
$N_{\pi\pi}=1.1\times N_4=5.9\pm2.5$, where 1.1 is the ratio of 
the total number  of $\pi\pi\gamma$ events to those in the  
$0.5<M_{\pi\pi}<1$~GeV/$c^2$  mass region and
is taken from simulation.
The expected $M_{p\bar{p}}$ spectrum for $\pi\pi\gamma$ events passing 
the $p\bar{p}\gamma$ selection criteria is shown 
as the dotted histogram in Fig.~\ref{pipi_m}.
 
The procedure used to estimate the background from the  
$e^+e^-\to K^+K^-\gamma$ process is similar to that used
to estimate the   $e^+e^-\to\pi^+\pi^-\gamma$ background.
The number of events in the $\phi$ meson peak in the 
distribution of
invariant mass of the charged particles
calculated under the kaon hypothesis is
used to determine the number of kaon events.
The total number of $KK\gamma$ events remaining after the standard 
selection criteria is 
estimated to be $N_{KK}=2.5\pm1.0$. The expected $M_{p\bar{p}}$ 
distribution for these events is shown 
as the solid histogram in Fig.~\ref{pipi_m}.

To estimate the electron background, the kinematic
properties of the $e^+e^-\to e^+ e^-\gamma$ process are used.
About 60\% of $e^+ e^-\gamma$ events have
$e^+e^-$ invariant mass between 3 and 7~GeV/$c^2$ and
$\cos{\psi^\ast}< -0.97$, where  $\psi^\ast$ is the
angle between the two tracks in the $e^+e^-$  center-of-mass frame.
In the event sample with two proton candidates, only one
event has the above characteristics.
With this event assumed to be background from $e^+e^-\to e^+ e^-\gamma$,
the total $e^+e^-\gamma$ background is estimated to be
$1.8\pm 1.8$ (0.8 events with $M_{p\bar{p}}<4.5$~GeV/$c^2$).

The method used to estimate $\mu^+\mu^-\gamma$ background relies on
the difference between the two-proton and the two-muon mass spectra.
From the simulation 44\% of $\mu^+\mu^-\gamma$ events are expected
to have a two-proton invariant mass greater than 4.5~GeV.
In data, only four such events are found, with an expected background
of $5\pm3$ events from the $e^+e^-\to p\bar{p}\pi^0$ process (see
Sec.~\ref{pppi}). From these numbers,  the
total muon background is estimated to not exceed 11 events.
A similar limit is obtained directly from $\mu^+\mu^-\gamma$
Monte Carlo simulation.
From about 2 million simulated $\mu^+\mu^-\gamma$ events
(20\% of the number of events expected in data), no events 
pass the  $p\bar{p}\gamma$ selection criteria, leading to a 90\% CL upper
limit of 12 events. The expected $M_{p\bar{p}}$ spectrum for
$e^+e^- \to \mu^+\mu^-\gamma$ events normalized to a total of 
11 events is shown as the dashed histogram
in Fig.~\ref{pipi_m}. This upper limit on the number of
muon events is used as a measure of the systematic uncertainty due to
$\mu^+\mu^-\gamma$ background. This uncertainty is calculated as 
a function of the $p\bar{p}$ mass and is added to the systematic error on
the number of $p\bar{p}\gamma$ events.

\subsection{$e^+e^-\to p\bar{p}\pi^0$ background \label{pppi}}
A dominant source of background to the $e^+e^-\to p\bar{p} \gamma$ process
arises from  $e^+e^-\to p\bar{p}\pi^0$.  A significant fraction 
of $p\bar{p}\pi^0$ events with an undetected low-energy
photon or with merged photons from the $\pi^0$ decay are reconstructed
under the $p\bar{p}\gamma$ hypothesis with a low value of $\chi^2$ and thus are
not easily separable from the process under study. 
Experimental data is used to devise a  procedure to subtract
this background.

\begin{figure}
\psfrag{pp}[][][0.7]{$\mbox{p}\bar{\mbox{p}}$}
\includegraphics[width=.48\linewidth]{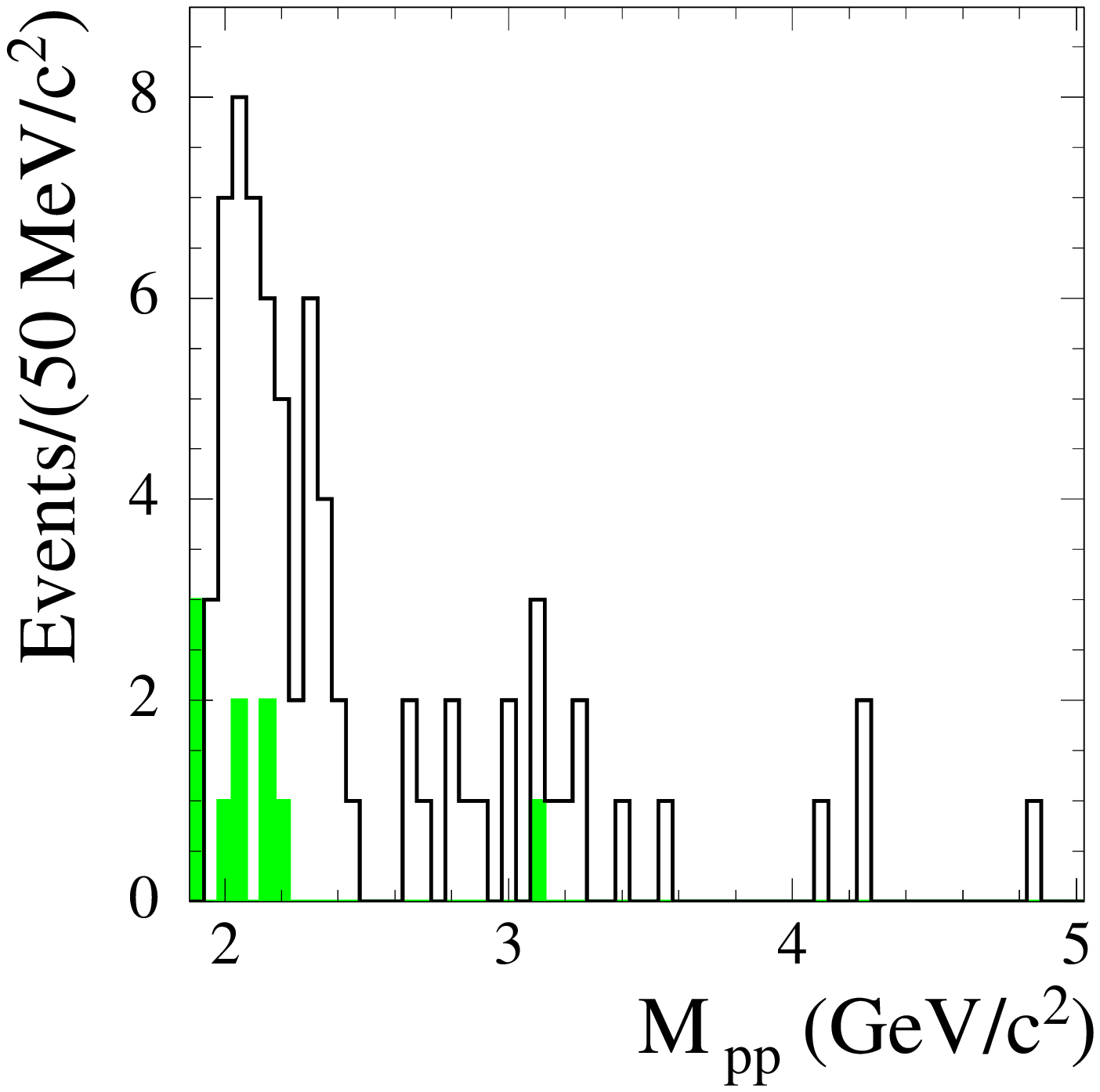}
\hfill
\includegraphics[width=.48\linewidth]{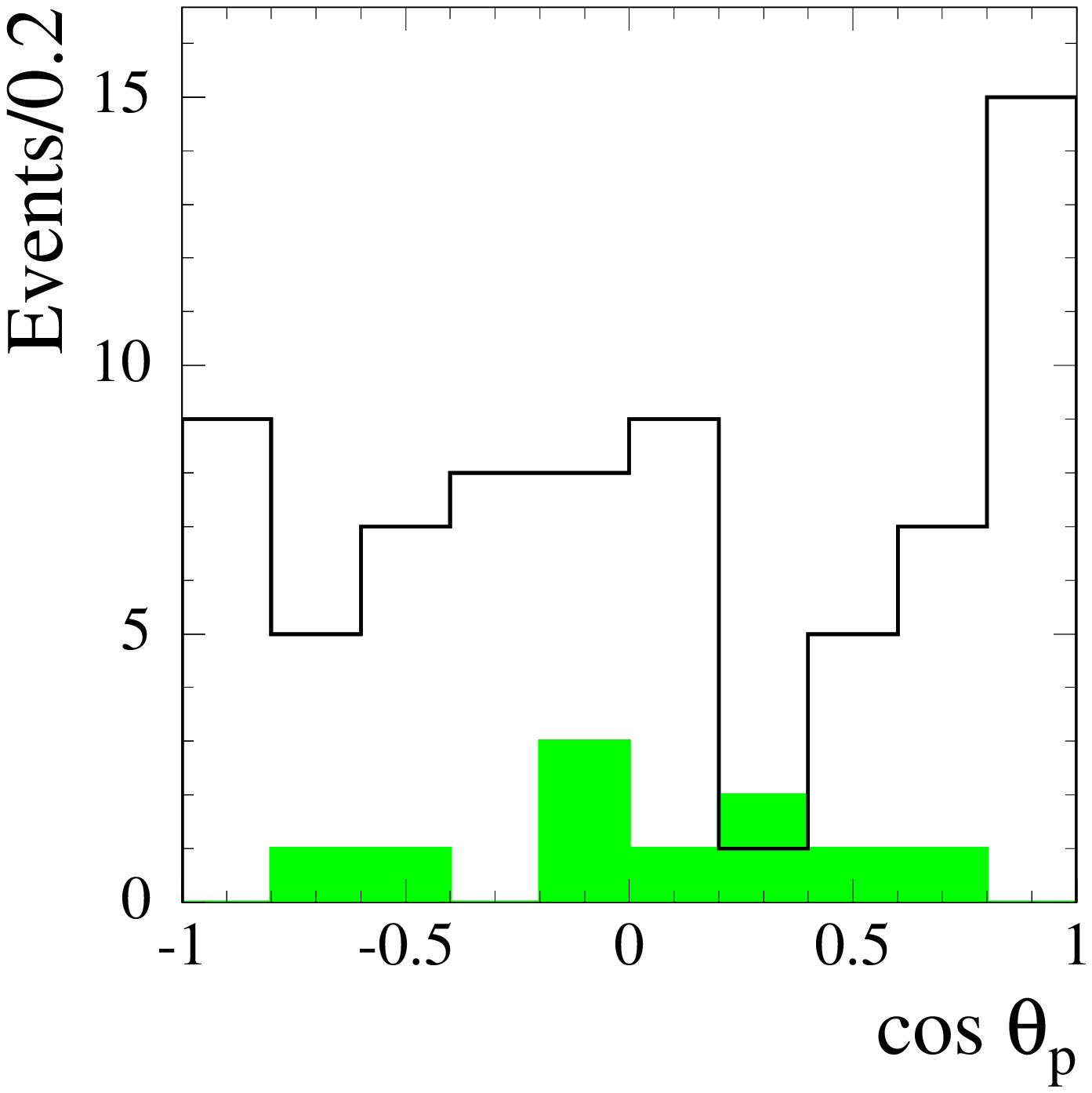}
\caption{The $M_{p\bar{p}}$  spectrum (left) and
the $\cos{\theta_{p}}$ distribution (right) for selected
$e^+e^-\to p\bar{p}\pi^0$ candidates in data. The shaded histogram shows the
background contribution estimated from $M_{\gamma\gamma}$ sidebands.
\label{pppi0_mpp}}
\end{figure}

For the $p\bar{p}\pi^0$ background study, events
with two charged particles identified as protons
and at least two photons with energy greater than 0.1~GeV,  one of
which must have center-of-mass energy above 3~GeV, are selected.
   The two-photon invariant mass is required to be in the
range 0.07 to 0.2~GeV/$c^2$.
A kinematic fit under the
$e^+e^-\to p\bar{p}\gamma\gamma$ hypothesis is then performed.
For events with more than two photons, all two-photon
combinations are analyzed and only the combination with the smallest
$\chi^2$ in the kinematic fit  is considered.
Requirements on the $\chi^2$ of the kinematic fit ($\chi^2<25$)
and the two-photon invariant mass ($0.1025<M_{\gamma\gamma}<0.1675$~GeV/$c^2$)
are then imposed on the  $e^+e^-\to p\bar{p}\pi^0$ candidates.
The sidebands
$0.0700<M_{\gamma\gamma}<0.1025$~GeV/$c^2$ and
$0.1675<M_{\gamma\gamma}<0.2000$~GeV/$c^2$ are used to
estimate background.
The $M_{p\bar{p}}$  spectra and $\cos{\theta_{p}}$ distributions
for data events from the signal and sideband regions are shown in
Fig.~\ref{pppi0_mpp}.
The total number of selected events is 74 in the signal region and
10 in the sidebands. The number of $e^+e^-\to p\bar{p}\pi^0$ events 
in the sidebands expected from MC simulation is 2.7.

\begin{figure}
\psfrag{pp}{$\mbox{p}\bar{\mbox{p}}$}
\includegraphics[width=.45\textwidth]{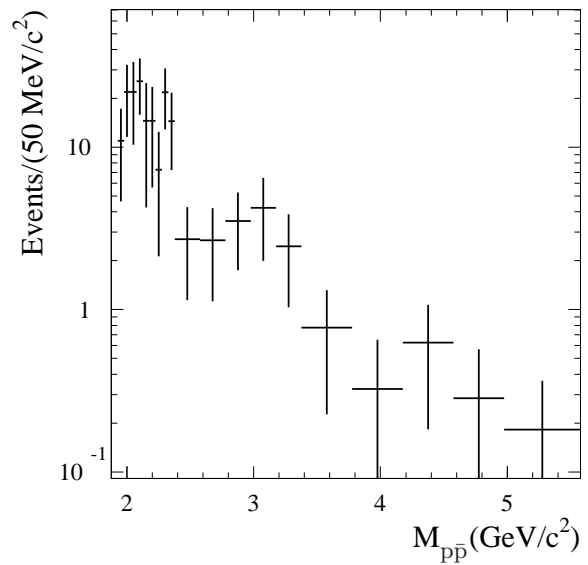}
\caption{The expected $M_{p\bar{p}}$ spectrum for
$e^+e^-\to p\bar{p}\pi^0$ events selected with the 
standard $p\bar{p}\gamma$
criteria. The spectrum is obtained by scaling the data distribution
shown in Fig.~\ref{pppi0_mpp}    
by the factor $K_{\mbox{MC}}(M_{p\bar{p}})$ described in the text.
\label{pppi0_mpp1}}
\end{figure}

The $p\bar{p}\gamma\gamma$ selection criteria described above
are applied to simulated $e^+e^-\to q\bar{q}$ events generated with
the JETSET package.
The predicted number of $e^+e^-\to p\bar{p}\pi^0$ events is
$73\pm7$.
These events have an
enhancement in the $M_{p\bar{p}}$ distribution near $p\bar{p}$ threshold,
similar to that in data (Fig.~\ref{pppi0_mpp}),
but the angular distribution is peaked at $\cos{\theta_{p}}=\pm1$
and is not consistent with the nearly flat distribution found in data.
To study these events, simulated $e^+e^-\to p\bar{p}\pi^0$
events are generated according to three-body phase space with
an additional weight proportional to $(M_{p\bar{p}}-1.86)^{\frac{3}{2}}$
(to imitate the $M_{p\bar{p}}$ distribution observed in
data). The resulting $\cos{\theta_{p}}$  distribution is flat.
With these simulated events,   $K_{\mbox{MC}}(M_{p\bar{p}})$ is calculated
as the  ratio of the $M_{p\bar{p}}$ distributions for events selected with the
standard $p\bar{p}\gamma$ criteria to those with the $p\bar{p}\pi^0$
criteria as a  function of $M_{p\bar{p}}$.
The value of  the ratio $K_{\mbox{MC}}(M_{p\bar{p}})$ varies between
3.7 near $M_{p\bar{p}}$ threshold to
2.0 at 5~GeV/$c^2$.
The expected $M_{p\bar{p}}$ spectrum for $e^+e^-\to p\bar{p}\pi^0$
 background passing the $p\bar{p}\gamma$ selection criteria is
shown in Fig.~\ref{pppi0_mpp1} and
is evaluated as
$K_{\mbox{MC}}(M_{p\bar{p}}) \times(dN/dM_{p\bar{p}})_{data}$,
where $(dN/dM_{p\bar{p}})_{data}$ is the mass distribution for
$e^+e^-\to p\bar{p}\pi^0$ events obtained above (Fig.~\ref{pppi0_mpp}).
In Table~\ref{pppi0_tab}, the number of selected
$e^+e^-\to p\bar{p}\gamma$
candidates and the expected number of
$e^+e^-\to p\bar{p}\pi^0$ background events
for different $p\bar{p}$ mass ranges  are given.
  (The $p\bar{p}$ mass ranges near the $J/\psi$ and $\psi(2S)$ 
  resonances are excluded.)
The background contribution  grows from 5\% near 
$p\bar{p}$ threshold to 40\%
at $M_{p\bar{p}}\approx 4$~GeV/$c^2$. All observed $p\bar{p}\gamma$ candidates
with $M_{p\bar{p}}> 4.5$~GeV/$c^2$ are consistent with $p\bar{p}\pi^0$ 
background.

\begin{table*}
\caption{ The number of selected $p\bar{p}\gamma$
candidates, $N_{p\bar{p}\gamma}$, and the number of background events from
the $e^+e^-\to p\bar{p}\pi^0$ process, $N_{p\bar{p}\pi^0}$, for different
ranges of $M_{p\bar{p}}$.
The $p\bar{p}$ mass ranges near the $J/\psi$ and $\psi(2S)$ resonances are excluded.
\label{pppi0_tab}}
\begin{ruledtabular}
\begin{tabular}{lccccc}
$M_{p\bar{p}}$ (GeV/$c^2$) & $<2.50$  & 2.50--3.05 & 3.15--3.60 & 3.75--4.50 & $>4.5$ \\
$N_{p\bar{p}\gamma}$& 3166     &   322    &    37    &   20    &    4   \\
$N_{p\bar{p}\pi^0}$ &$171\pm29$&$33\pm11$ & $17\pm7$ & $8\pm4$ & $5\pm3$ \\
\end{tabular}
\end{ruledtabular}
\end{table*}

The JETSET simulation is used to find other possible sources
   of background from $e^+e^-\to q\bar{q}$.
The number of $q\bar{q}$ events with final states other 
than $p\bar{p}\pi^0$ passing all cuts is  $26\pm4$, with two final 
states, $p\bar{p}2\pi^0$ and $p\bar{p}\eta$,  accounting for 17 and 
5 events, respectively.  
The background contribution from these sources is estimated
from data using the $\chi^2$ sideband as described below.

\subsection{$e^+e^-\to p\bar{p}\pi^0\gamma$ background}

\begin{figure}
\psfrag{pp}[][][0.7]{$\mbox{p}\bar{\mbox{p}}$}
\includegraphics[width=.48\linewidth]{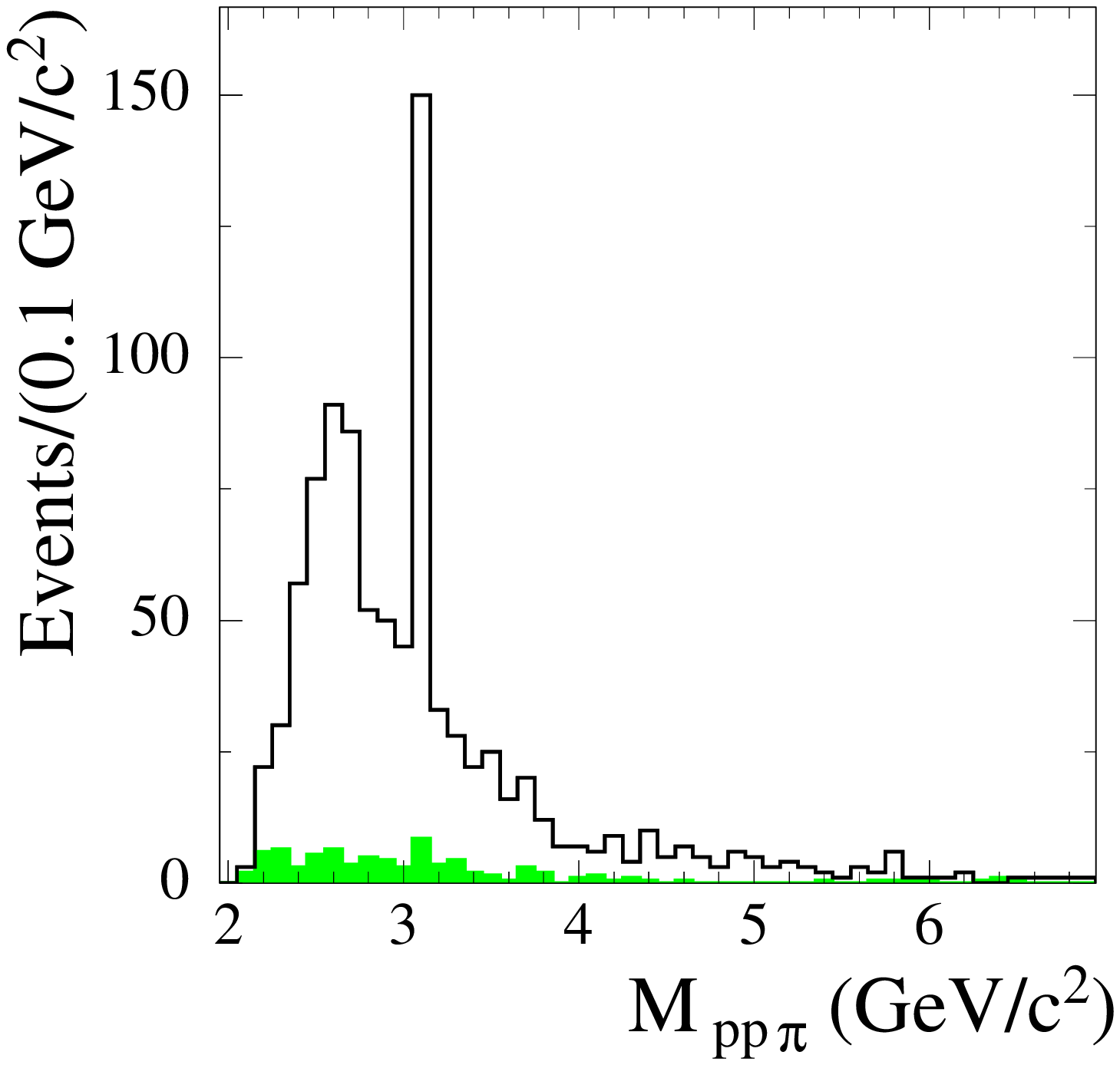}
\hfill
\includegraphics[width=.48\linewidth]{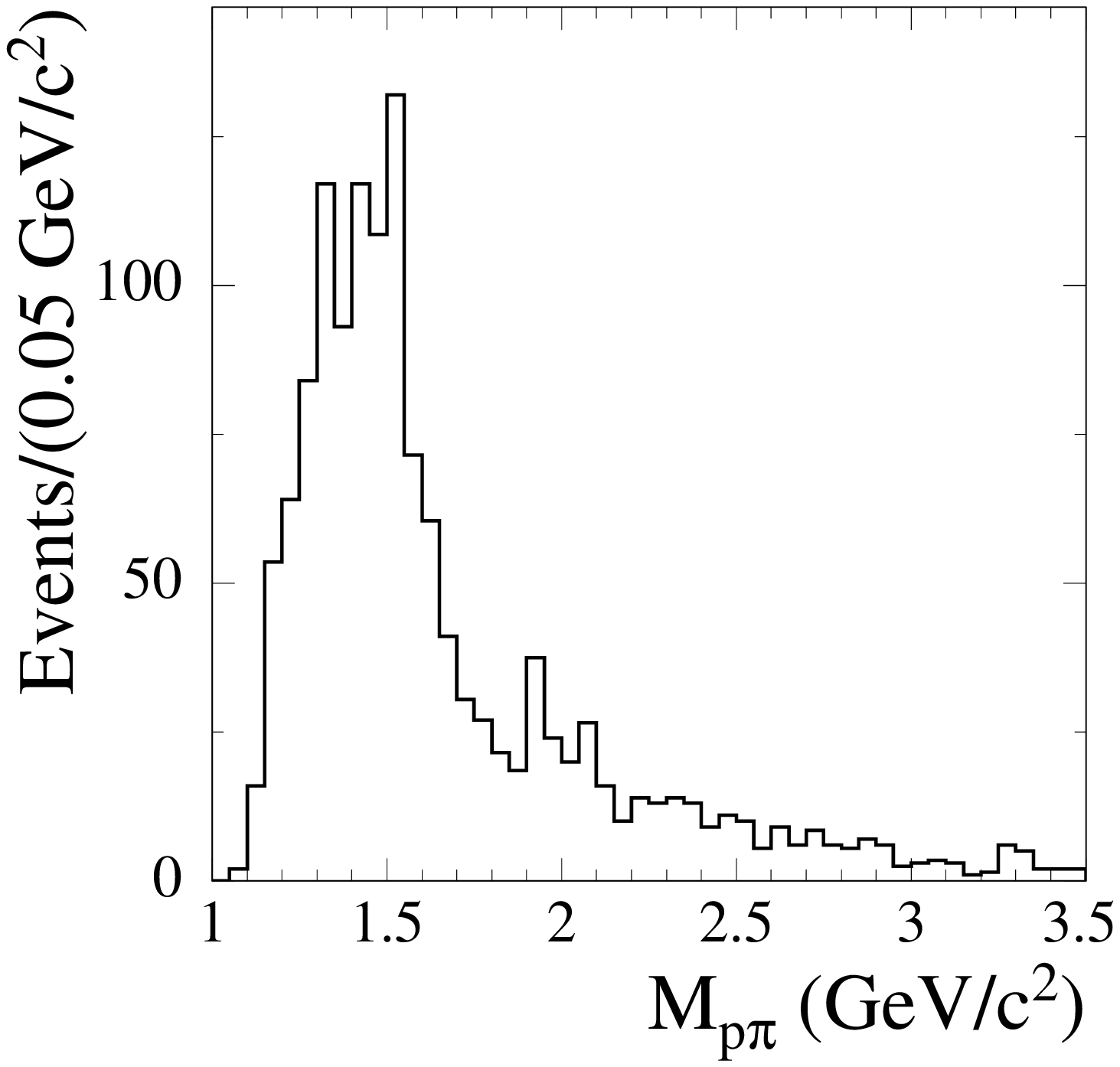}
\caption{Left: the $p\bar{p}\pi^0$ mass spectrum for
selected $e^+e^-\to p\bar{p}\pi^0\gamma$ candidates
with $\chi^2<20$. The hashed  histogram shows the background
contribution estimated from the sideband with $30<\chi^2<50$.
Right: the $p\pi^0$ ($\bar{p}\pi^0$) mass spectrum for
$e^+e^-\to p\bar{p}\pi^0\gamma$ data events with $p\bar{p}\pi^0$ mass
outside the $J/\psi$ peak.
\label{pppi0g_m}}
\end{figure}

The dominant ISR background process with protons in the final state
is $e^+e^-\to p\bar{p}\pi^0\gamma$.  To estimate this background,
events are selected with two charged particles identified as protons 
and at least three photons with energy greater than 0.1~GeV, with 
one of these photons having center-of-mass energy above 3~GeV.
The invariant mass of the two least energetic photons is required to be
   in the range 0.07-0.20~GeV/$c^2$. 
For events that pass these criteria, a kinematic fit
under the $e^+e^-\to p\bar{p}\pi^0\gamma$ hypothesis is performed.
The distribution of $p\bar{p}\pi^0$
invariant mass for events with $\chi^2<20$  is shown in 
Fig.~\ref{pppi0g_m}~(left). The shaded histogram shows the background
contribution estimated from the $\chi^2$ sideband: $30<\chi^2<50$.
Most  $e^+e^-\to p\bar{p}\pi^0\gamma$ events
have a $p\bar{p}\pi^0$ mass near a peak at 2.6~GeV. The contribution 
of  the $J/\psi\to p\bar{p}\pi^0$ decay is
also seen. The $p\pi^0$ ($\bar{p}\pi^0$) mass spectrum for
events with $p\bar{p}\pi^0$ mass away from the $J/\psi$ 
resonance is shown in
Fig.~\ref{pppi0g_m}~(right). The mass and width of the peak
dominating in this distribution agree with the parameters of
the $N(1440)$ state, suggesting that the main mechanism in the 
$e^+e^-\to p\bar{p}\pi^0$ reaction is a transition through
$N(1440)\bar{p}$ or $\bar{N}(1440)p$ intermediate states.

The number of $e^+e^-\to p\bar{p}\pi^0\gamma$ events passing the  
$p\bar{p}\gamma$ selection is estimated using
Monte Carlo simulation; $e^+e^-\to p\bar{p}\pi^0\gamma$ events
are generated in the $N(1440)\bar{p}\gamma+
\bar{N}(1440)p\gamma$ model with the $Np$
form factor reproducing the experimental $p\bar{p}\pi^0$ mass
      distribution. 
In the simulation, the ratio of detection
efficiencies for the $p\bar{p}\gamma$ and $p\bar{p}\pi^0\gamma$ 
selection criteria is $(1.5\pm0.2)\%$. From $847\pm31$  
selected $e^+e^-\to p\bar{p}\pi^0\gamma$
candidates (Fig.~\ref{pppi0g_m}) the background
contribution to the sample of $e^+e^-\to p\bar{p}\gamma$  candidates 
is estimated
to be $13\pm3$ events (about 0.3\% of the total number of
selected $p\bar{p}\gamma$ candidates).

The background contribution  from ISR processes with higher
multiplicity is significantly lower. A procedure similar to that 
described above is used to estimate the background from the 
$e^+e^-\to p\bar{p}2\pi^0\gamma$ process. Performing a kinematic 
fit under the $e^+e^-\to p\bar{p}2\pi^0\gamma$ hypothesis,  
$560\pm30$ events are selected. From the Monte Carlo simulation,
the ratio of  detection efficiencies for the
$p\bar{p}\gamma$ and $p\bar{p}2\pi^0\gamma$ selection criteria
is $(0.09\pm0.06)\%$, and the background contribution due to 
$e^+e^-\to p\bar{p}2\pi^0\gamma$ is estimated to be 
$0.5\pm0.3$ events.

\subsection{Background subtraction}\label{bkgsub}

\begin{table*}
\caption{$N_1$ and $N_2$ are the numbers of selected $p\bar{p}\gamma$ 
candidates with a
kinematic fit $\chi^2_{p}<30$ and $30<\chi^2_{p}<60$, respectively,
for signal and for different background processes. The last column 
shows the numbers of candidates selected in data. $\beta_i$ is the ratio
$N_2/N_1$ obtained from simulation. The numbers for 
$e^+e^-\to p\bar{p}\gamma$ are obtained from data using the background 
subtraction procedure described in the text.
\label{beta}}
\begin{ruledtabular}
\begin{tabular}{lccccccc}
&$\pi^+\pi^-\gamma$&$K^+K^-\gamma$&$p\bar{p}\pi^0$&
$p\bar{p}\pi^0\gamma$&$uds$&$p\bar{p}\gamma$&data\\
$N_1$ &$5.9\pm2.5$&$2.5\pm1.0$ &$229\pm32$&$13\pm3$&$26\pm4$&$3737\pm75$&4025\\
$\beta_i$&$0.71\pm0.05$&$0.52\pm0.04$&$0.13\pm0.01$&$1.53\pm0.25$
&$1.44\pm0.30$&$0.048\pm0.001$\\
$N_2$ & $4.2\pm1.8$ & $1.3\pm0.5$ &$29\pm5$ &$20\pm3$&$37\pm5$& $179\pm5$&288\\
\end{tabular}
\end{ruledtabular}
\end{table*}

Table~\ref{beta} summarizes the expected number of background events estimated
in the above sections.
The $uds$ column shows the number of background events expected from
$e^+e^-\to q\bar{q}$ with the $p\bar{p}\pi^0$ final
  state excluded. This background is estimated using the JETSET event generator.
Because JETSET has not been precisely verified 
for the rare processes contributing to the $p\bar{p}\gamma$
candidate sample,  the background estimation is 
based on the difference in $\chi^2$ distributions
for signal and background events. The second row in Table~\ref{beta}
lists $\beta_i$, the ratio of $N_2$, the number of events with
     $30<\chi^2_{p}<60$,  to $N_1$, the number of events with $\chi^2_{p}<30$,
calculated for signal and background processes using 
the Monte Carlo simulation.
The last row  in Table~\ref{beta} shows 
the expected numbers of signal and background events
in the $\chi^2$ sideband ($30<\chi^2_{p}<60$) calculated
as $N_2=\beta N_1$. 
In the Table, it is evident that $\chi^2$ distributions
for signal events and those for background from the
processes with higher hadron multiplicity 
(columns labeled $uds$ and $p\bar{p}\pi^0\gamma$) are very different. 
This difference can be used
to estimate the background from these two sources, as follows.
First,  the $\pi^+\pi^-\gamma$, $K^+K^-\gamma$, $e^+e^-\gamma$, and
$p\bar{p}\pi^0$ background determined in previous sections is subtracted 
from data.
Then, from the resulting numbers of events in the signal and sideband
$\chi^2$ regions, $N_1^\prime$ and $N_2^\prime$, the 
numbers of signal and background (from $uds$ and $p\bar{p}\pi^0\gamma$
     sources) events with $\chi^2_{p}<30$ can be calculated:
\begin{equation}
N_{sig}=\frac{N_1^\prime-N_2^\prime/\beta_{bkg}}
{1-\beta_{p\bar{p}\gamma}/\beta_{bkg}},\; N_{bkg}=N_1^\prime-N_{sig},
\label{bkgsub_eq}
\end{equation}
where $\beta_{bkg}$ is the ratio of fractions of events 
in the sideband and signal $\chi^2$ regions
averaged over $uds$ and $p\bar{p}\pi^0\gamma$ backgrounds.
For this coefficient $\beta_{bkg}=1.5\pm0.4$ is used;
it is the  average of
$\beta_{uds}$ and $\beta_{p\bar{p}\pi^0\gamma}$ with an uncertainty covering
the full range of $\beta_{uds}$ and $\beta_{p\bar{p}\pi^0\gamma}$ variations.

        In Table~\ref{beta},  it is also evident
that $p\bar{p}\gamma$ events dominate the sideband.
Therefore, the  background is very sensitive
to the accuracy of the $\beta_{p\bar{p}\gamma}$ coefficient. In particular, the
data-Monte Carlo difference in the $\chi^2$ distribution can lead to a systematic
shift of the result. The simulation of the $\chi^2$ distribution
for $p\bar{p}\gamma$ events is validated using data and simulated events in 
the channels $e^+e^-\to \mu^+\mu^-\gamma$ and $e^+e^-\to K^+K^-\gamma$,
both of which are
kinematically very similar to the process under study. 
In the simulations, 
the $\beta$ coefficients for all three processes agree within 2\%.
The ratio the $\beta $ coefficients for data and simulation  is 
$1.01\pm0.03$ for $e^+e^-\to K^+K^-\gamma$ and
$1.015\pm0.012$ for $e^+e^-\to \mu^+\mu^-\gamma$. The 
  $\mu^+\mu^-\gamma$ ratio is used to correct the $\beta_{p\bar{p}\gamma}$ value
obtained from simulation, which results in 
$\beta_{p\bar{p}\gamma}=0.048\pm0.003$.
The error is estimated using the $\beta_{p\bar{p}\gamma}$ variation 
as a function of $p\bar{p}$ mass.

With the method described above,  the total number
of $e^+e^-\to p\bar{p}\gamma$ events ($N_{sig}$) and 
background events from $uds$ and $p\bar{p}\pi^0\gamma$ sources ($N_{bkg}$)
in the signal region are $3737 \pm 67 \pm 34$ and 
$50\pm12\pm16$, respectively.
The main source of the systematic uncertainty on $N_{sig}$ is
the uncertainty in the $p\bar{p}\pi^0$ background. The  numbers
of $uds$ and $p\bar{p}\pi^0\gamma$ background events
are in good agreement with their estimations from simulation,
$(13\pm3) + (26\pm4)=39\pm5$. The total background in the
$\chi^2_{p}<30$ region is $288$ events, about 8\%
of the number of signal events.  

The background subtraction procedure is performed in each $p\bar{p}$  mass bin.
The resulting numbers of signal events for each bin are listed in
Table~\ref{sumtab}. 
The events from $J/\psi$ and $\psi(2S)$ decays are subtracted from the
contents of the corresponding bins (see Sec.~\ref{jpsi}).

\section{Angular distribution}
The ratio of electric and magnetic
form factors is extracted by analysing the distribution 
of $\theta_p$, the angle between the proton momentum in
the $p\bar{p}$ rest frame and the momentum of the $p\bar{p}$ system
in the $e^+e^-$ center-of-mass frame. 
In general, this distribution is given by
\begin{eqnarray}
\lefteqn{\frac{{\rm d}N}{{\rm d}\cos{\theta_p}}=} \nonumber \\
&A\left(H_M(\cos{\theta_p},M_{p\bar{p}})+
\left |\frac{G_E}{G_M}\right|^2H_E(\cos{\theta_p},M_{p\bar{p}})\right).
\label{an_fit}
\end{eqnarray}
The functions $H_M(\cos{\theta_p},M_{p\bar{p}})$ and 
$H_E(\cos{\theta_p},M_{p\bar{p}})$
do not have simple analytic forms, 
but are determined using MC simulation.
Two samples of $e^+e^-\to p\bar{p}\gamma$ events are generated,
one with $G_E=0$ and the other with $G_M=0$. 
The obtained functions are 
similar to the  $1+\cos^2 \theta_p$ and $\sin^{2}\theta_p$ functions
describing angular distributions for magnetic and electric form factors
in the case of $e^+e^-\to p\bar{p}$ process.

\begin{figure}
\includegraphics[width=0.48\linewidth]{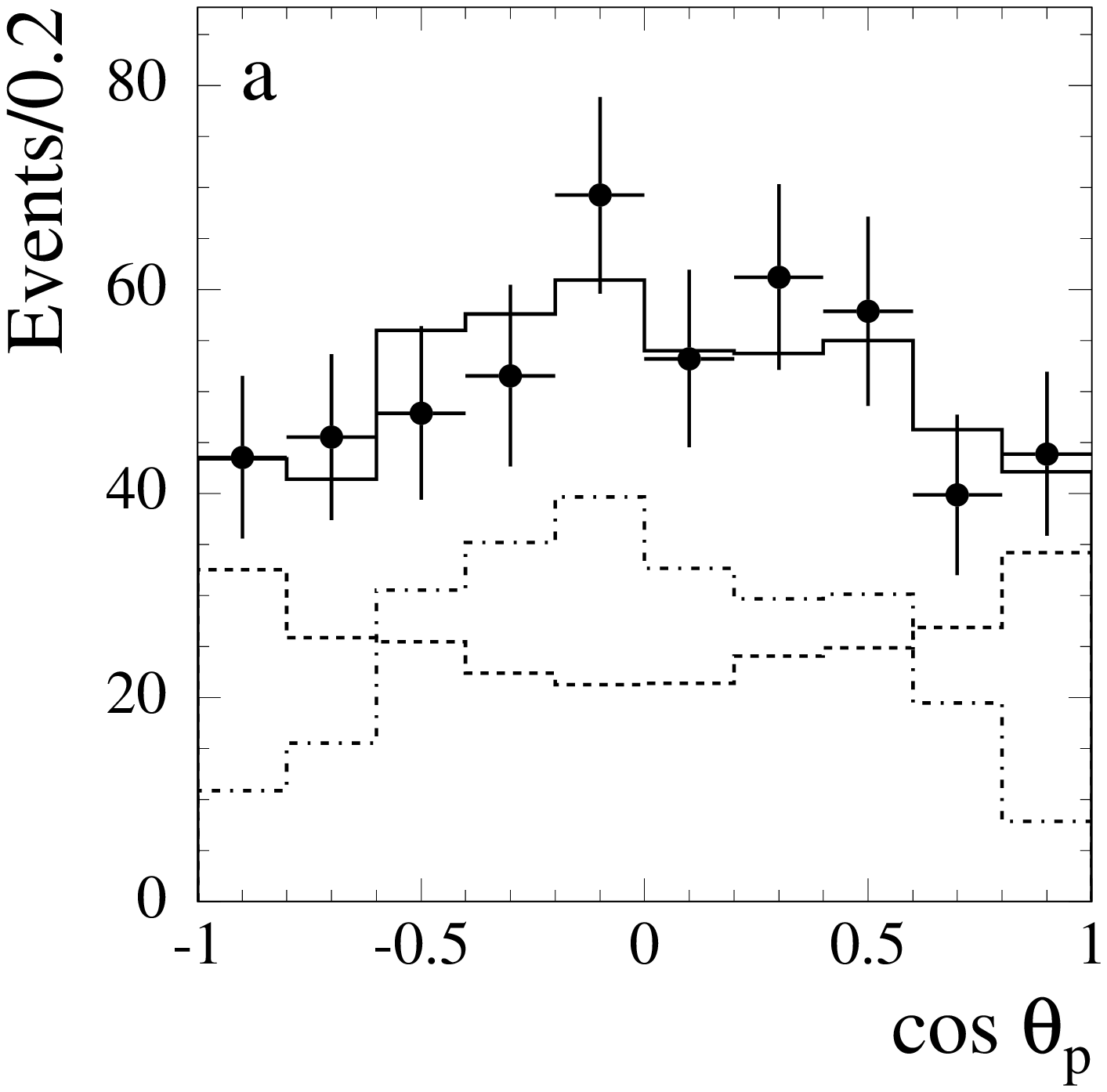}
\hfill
\includegraphics[width=0.48\linewidth]{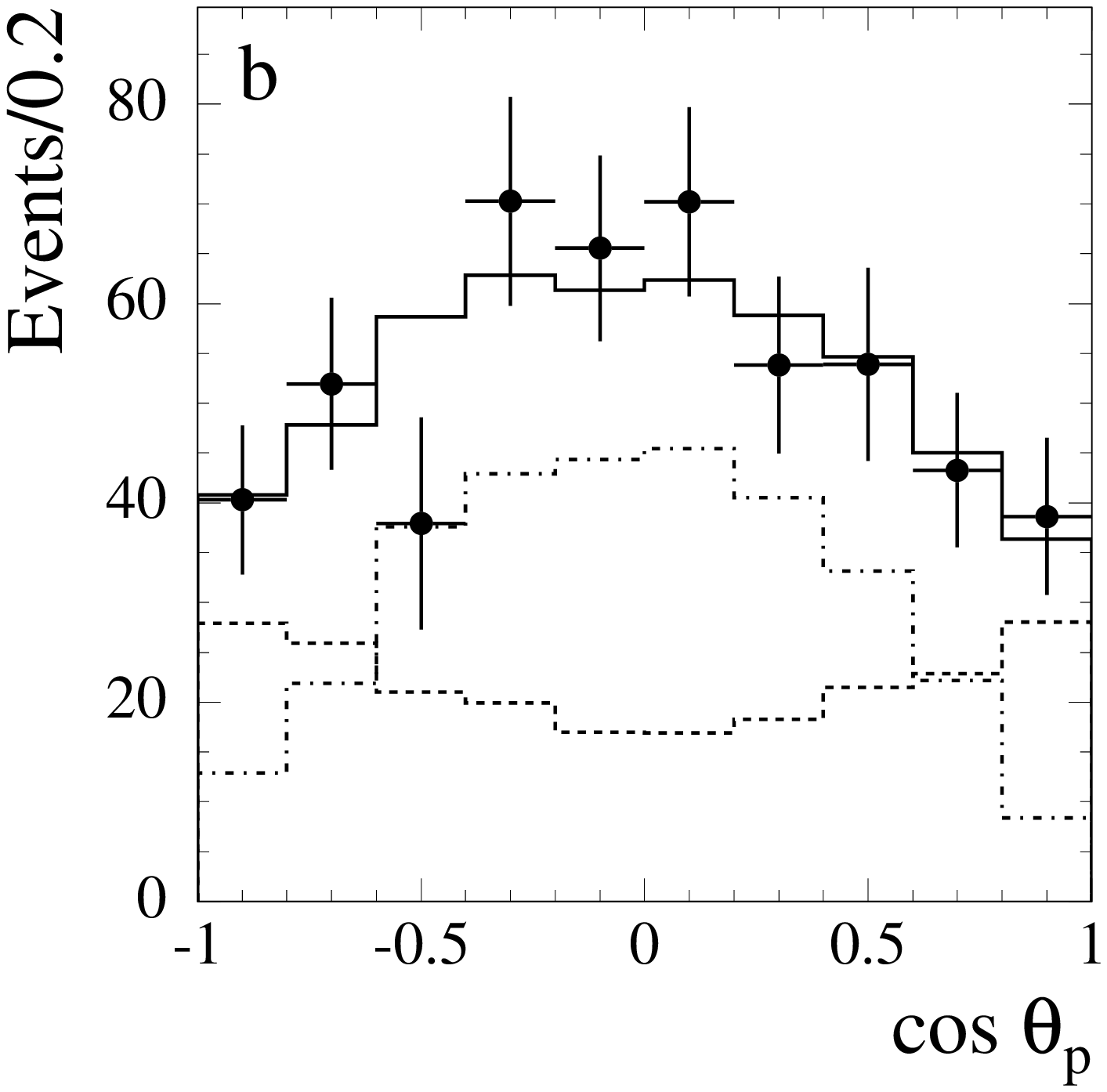}
\vspace{2mm}\\
\includegraphics[width=0.48\linewidth]{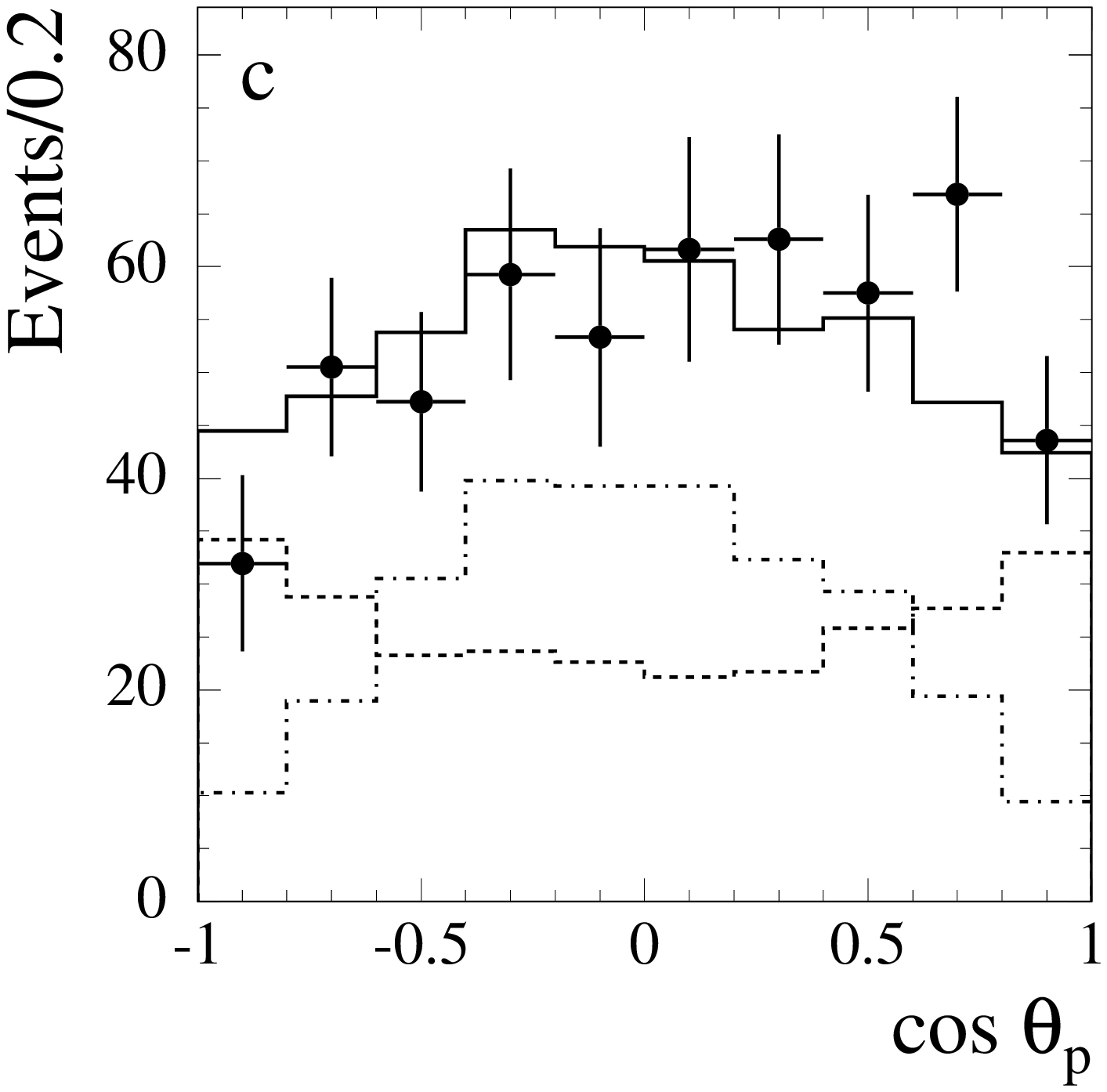}
\hfill
\includegraphics[width=0.48\linewidth]{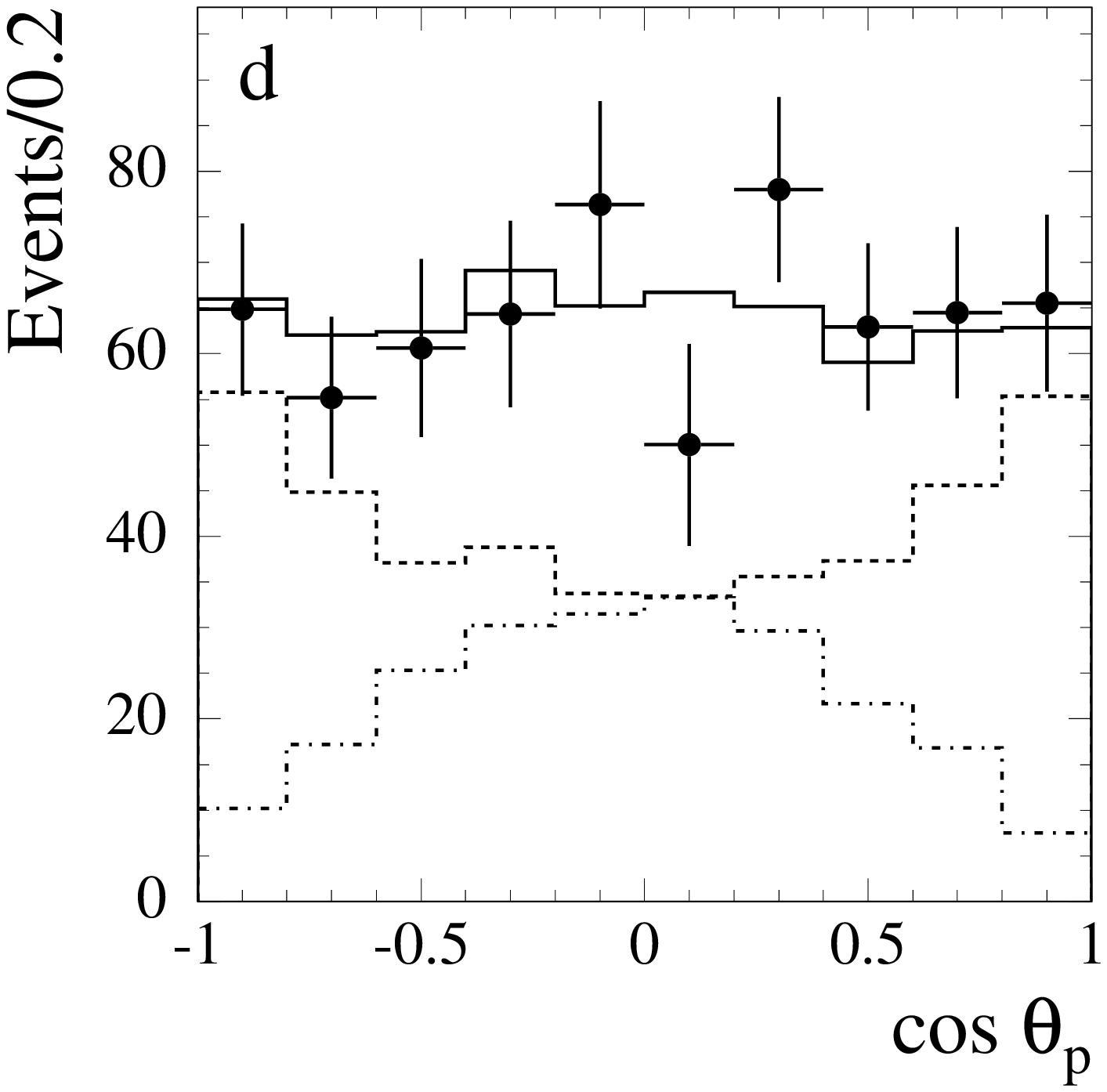}
\vspace{2mm}\\
\includegraphics[width=0.48\linewidth]{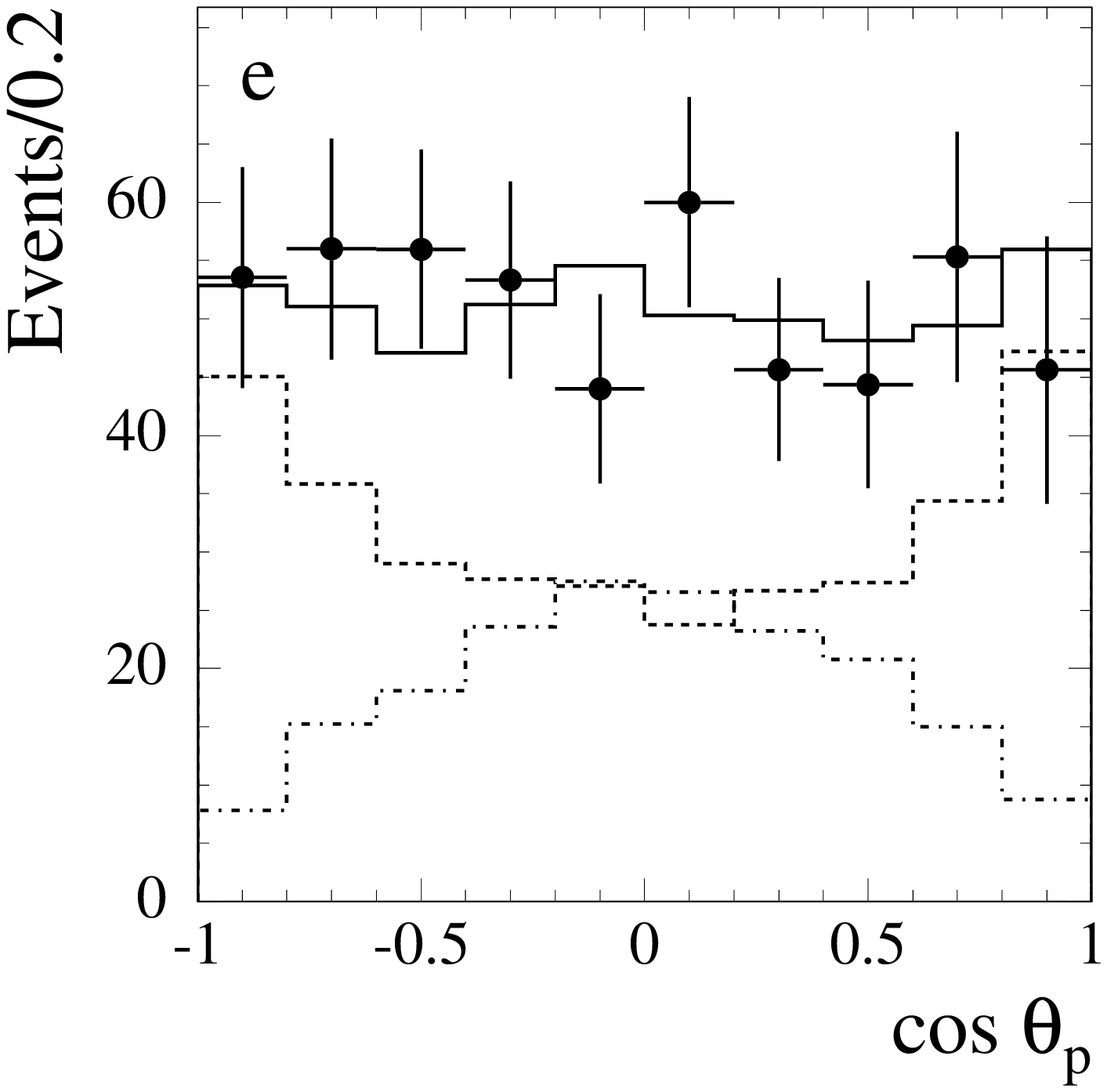}
\hfill
\includegraphics[width=0.48\linewidth]{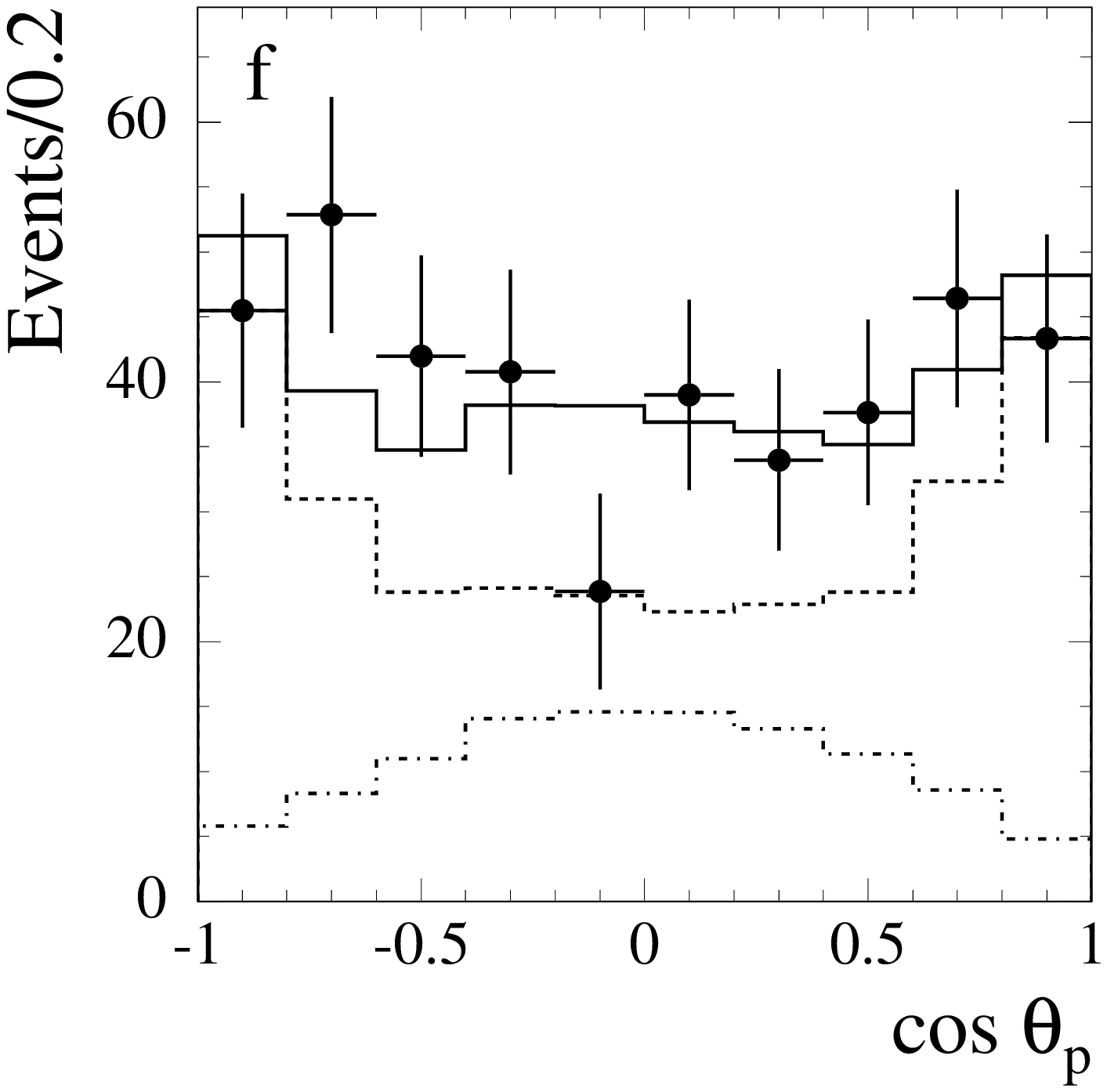}
\caption{The $\cos{\theta_p}$ distributions for
different $p \bar{p}$ mass regions:
(a) 1.877--1.950~GeV/$c^2$,
(b) 1.950--2.025~GeV/$c^2$,
(c) 2.025--2.100~GeV/$c^2$,
(d) 2.100--2.200~GeV/$c^2$,
(e) 2.200--2.400~GeV/$c^2$,
(f) 2.400--3.000~GeV/$c^2$.
The points with error bars show data distributions after
background subtraction.  
The histograms are fit results: the dashed histograms
show the contributions  corresponding
to the magnetic form factor; the dash-dotted histograms
show the contributions from the electric
      form factor.
\label{as}}
\end{figure}

The angular distributions of the data are fit in 
six ranges of 
$p \bar{p}$ invariant mass
from threshold to 3~GeV/$c^2$.
to measure $|{G_E}/{G_M}|$. 
The fit intervals, the corresponding numbers of selected
events, and  the estimated numbers of background events are listed
in Table~\ref{ang_tab}.
\begin{table}
\caption{$N$ is the number of selected $p\bar{p}\gamma$
candidates, $N_{bkg}$ is the number of background events, and
$|G_E/G_M|$ is the fitted ratio of form factors,  for
each $p\bar{p}$ mass interval.\label{ang_tab}}
\begin{ruledtabular}
\begin{tabular}{cccc}
$M_{p\bar{p}}$, GeV/$c^2$ & $N$  & $N_{bkg}$ & $|G_E/G_M|$ \\ 
1.877--1.950        & 533  & $ 2\pm7 $ & $1.41_{-0.22-0.12}^{+0.24+0.17}$ \\
1.950--2.025        & 584  & $37\pm12$ & $1.78_{-0.25-0.14}^{+0.31+0.18}$ \\
2.025--2.100        & 602  & $50\pm15$ & $1.52_{-0.23-0.12}^{+0.27+0.16}$ \\
2.100--2.200        & 705  & $42\pm14$ & $1.18_{-0.19-0.11}^{+0.20+0.12}$ \\
2.200--2.400        & 592  & $61\pm16$ & $1.32_{-0.23-0.14}^{+0.26+0.17}$ \\
2.400--3.000        & 464  & $45\pm12$ & $1.22_{-0.30-0.16}^{+0.30+0.16}$ \\
\end{tabular}
\end{ruledtabular}
\end{table}
For each $p\bar{p}$ mass interval and  each angular bin the background is subtracted 
using the procedure described in Section \ref{bkgsub}. 
The
angular distributions obtained are shown in Fig.~\ref{as}.
The distributions are fit to Eq.~\ref{an_fit} with
two free parameters $A$ (the overall normalization)
and $|G_E/G_M|$. The
functions $H_M$ and $H_E$ are modeled with the histograms
obtained from MC simulation with the $p\bar{p}\gamma$ selection
applied.
To account for differences between the $p\bar{p}$ mass distributions
of $p\bar{p}\gamma$ events in data and  MC simulation,
the histograms $H_M$ and $H_E$ are re-calculated using weighted
events.
The weights are obtained from the ratio
of the $p\bar{p}$ mass distributions in data and simulation.
In principle,  the weights for $H_M$ and $H_E$ differ
due to the different mass dependences of $G_M$ and $G_E$.  
A first approximation uses $G_M=G_E$. The fitted values of
$|G_E/G_M|$ are then used in the next approximation
to recalculate $H_M$ and $H_E$. The second iteration
leads to a small change (less than 2\%) of the fitted values,
and the procedure converges after a third iteration.

\begin{figure}[h]
\includegraphics[width=.45\textwidth]{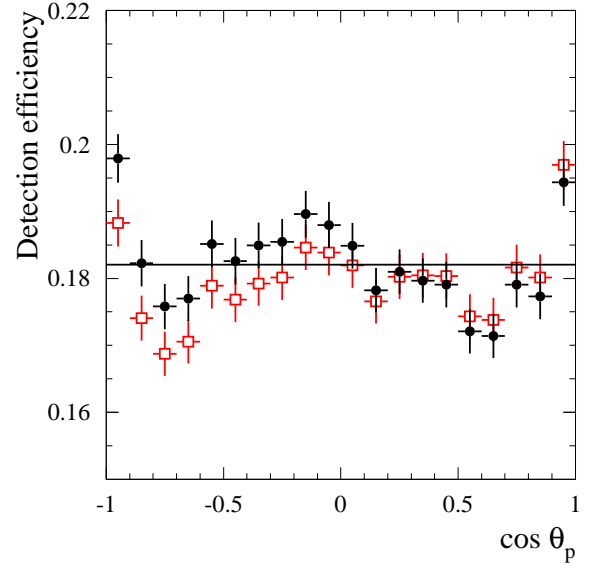}
\caption{The angular dependence of the detection efficiency
for simulated events with $M_{p\bar{p}}<2.5$~GeV/$c^2$
before (open squares) and after (filled circles) correction
for data-simulation difference in detector response.
\label{effvsang}}
\end{figure}
The simulated angular distributions are corrected to
account for the differences between the data and the simulations, in
particle identification, tracking, and photon efficiencies. These corrections
are discussed in detail in the next section.
The angular dependences of detection efficiencies calculated
with MC simulation before and after corrections
are shown in Fig.~\ref{effvsang}.
The variations from uniform, which do not exceed 10\%
fractionally,
derive from the momentum dependences of proton/antiproton 
particle identification efficiencies.
These manifest themselves as angular variations because
there is a
strong correlation between
proton/antiproton momentum and $\theta_p$.
In particular, the minima in detection efficiency at 
$| \cos{\theta_p} |= 0.75$ correspond to the minima in
proton/antiproton identification efficiencies
for momenta near 1.5~GeV/$c$.

\begin{figure}
\psfrag{pp}{$\mbox{p}\bar{\mbox{p}}$}
\includegraphics[width=.45\textwidth]{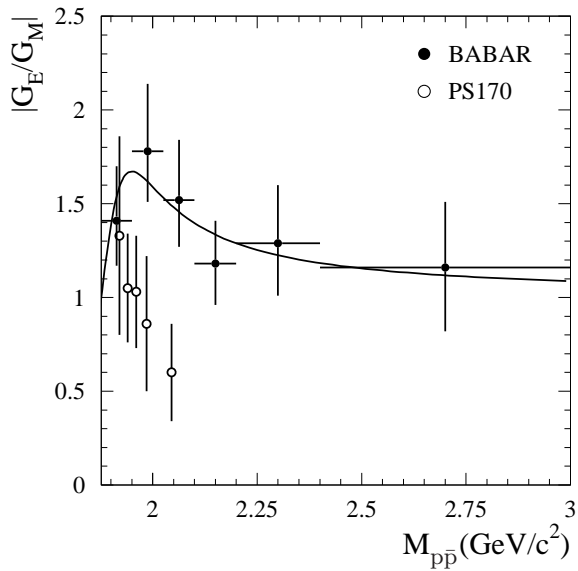}
\caption{ The measured $|G_E/G_M|$ mass dependence.
Filled circles depict \babar\ data, the curve is the fit result.
Open circles show the data from PS170~\cite{LEAR}.
\label{gegm}}
\end{figure}

The histograms fit to the angular
distributions are shown in Fig.~\ref{as}; the values of
$|G_E/G_M|$ are listed in Table~\ref{ang_tab} and shown in
Fig.~\ref{gegm}. The curve in Fig.~\ref{gegm}
($1+ax/(1+bx^2)$) is used
in the iteration procedure to calculate the weight.
The quoted errors on $|G_E/G_M|$ are statistical and systematic.
The dominant contribution to the systematic error comes from
the uncertainty in the $p\bar{p}\pi^0$ background. For example,
 for the 1.950-2.025~GeV/$c^2$ range this contribution to the lower (upper) error
is 0.12 (0.16), which dominates the total systematic error in 
this bin, and likewise dominates systematic error in all bins.
The error due to the limited MC simulation
statistics (0.08 for 1.950-2.025~GeV/$c^2$),
uncertainties in the coefficients $\beta$ used for background
subtraction (0.01), the uncertainty of description of
mass dependence of $|G_E/G_M|$ (0.01), and the  uncertainty
in the efficiency correction (0.02) are all considered.
 The last is  conservatively estimated
  as the difference between fitted values of $|G_E/G_M|$ obtained
with and without applying the efficiency correction.

\begin{figure}
\includegraphics[width=.45\textwidth]{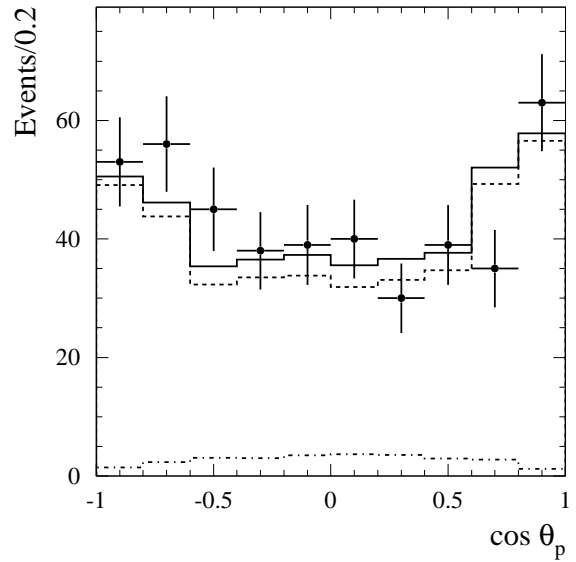}
\caption{The $\cos{\theta_p}$ distributions for
$J/\psi\to p\bar{p}$ decay. The points with error bars
correspond to the background-subtracted data distribution, the solid histogram 
is the fit result, and the dashed and dashed-dotted histograms
show the fit contributions from the magnetic and electric form factors,
respectively. 
\label{aspsi}}
\end{figure}
The angular distribution for $J/\psi\to p\bar{p}$
decay has also been studied. Its shape is  
commonly parameterized using
the form $1+\alpha \cos^2{\vartheta}$. 
The 
coefficient $\alpha$ has been measured with relatively high precision 
in several experiments~\cite{psipp1,BESpp}, and its average value 
     is $\alpha= 0.660\pm0.045$.
The \babar\ data distribution for
$J/\psi\to p\bar{p}$ decay  is shown in Fig.~\ref{aspsi}. 
The non-peaking background is subtracted by taking 
the difference between
the histograms for the signal mass region (3.05-3.15~GeV/$c^2$) and
the mass sidebands (3.00--3.05 and 3.15--3.20~GeV/$c^2$).
The fitting procedure used is similar to the one described above with
$\alpha=(1-g\tau)/(1+g\tau)$, where $g=|G_E/G_M|^2$ and
$\tau=4m_p^2/M_{J/\psi}^2$. 
The resulting value 
$\alpha=0.75^{+0.42}_{-0.35}$ is in agreement with the world average  
but has significantly larger uncertainty.

\begin{figure}
\includegraphics[width=.45\textwidth]{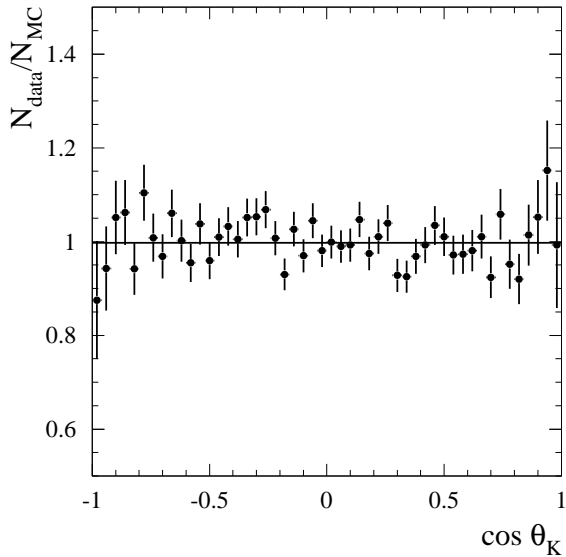}
\caption{The ratio of data and simulated distributions
of $\cos{\theta{_K}}$ for $e^+e^-\to K^+K^-\gamma$ process.
\label{askk1}}
\end{figure}

To cross-check this method to measure $|G_E/G_M|$,
a comparison is made between  the data and simulated distributions of
$\cos{\theta_{K}}$ for $e^+e^-\to\phi\gamma\to K^+K^-\gamma$ process.
Here $\theta_K$ is defined analogous to the definition of  $\theta_p$: 
$\theta_K$ is the angle between the 
$K^-$ momentum in the $K^+K^-$ rest frame  and the momentum of the 
$K^+K^-$ system in the $e^+e^-$ center-of-mass frame. 
The angular dependence for this process is well known
 (approximately  $\cos^2 \theta_K$) and
event kinematics are similar to the $e^+e^-\to p\bar{p}\gamma$
kinematics near threshold.  Figure~\ref{askk1} shows the ratio of 
data and simulated distributions over $\cos \theta_K$ 
for events with the $K^+K^-$ mass near the $\phi$. 
The simulation describes the angular dependence of detection
efficiency well.

When these \babar\ measurements of the $|G_E/G_M|$ ratio are compared with
the PS170 measurements~\cite{LEAR}  (Fig.~\ref{gegm}),
a large disagreement is seen for $M_{p\bar{p}}$ 
larger than 1.93~GeV/$c^2$.

\section{Detection efficiency}\label{sdetef}
\begin{figure}
\psfrag{pp}[][][0.7]{$\mbox{p}\bar{\mbox{p}}$}
\includegraphics[width=.48\linewidth]{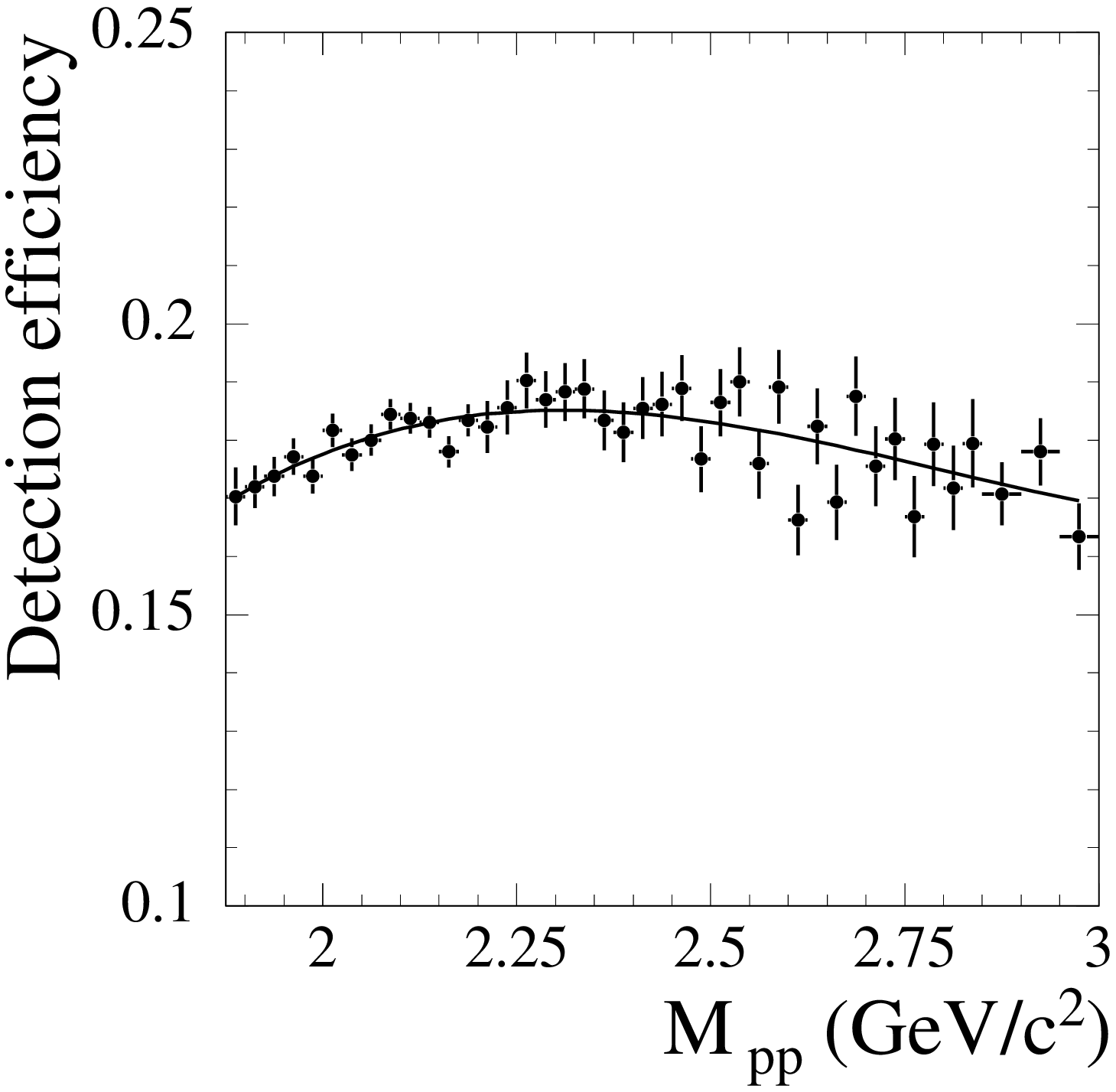}
\hfill
\includegraphics[width=.48\linewidth]{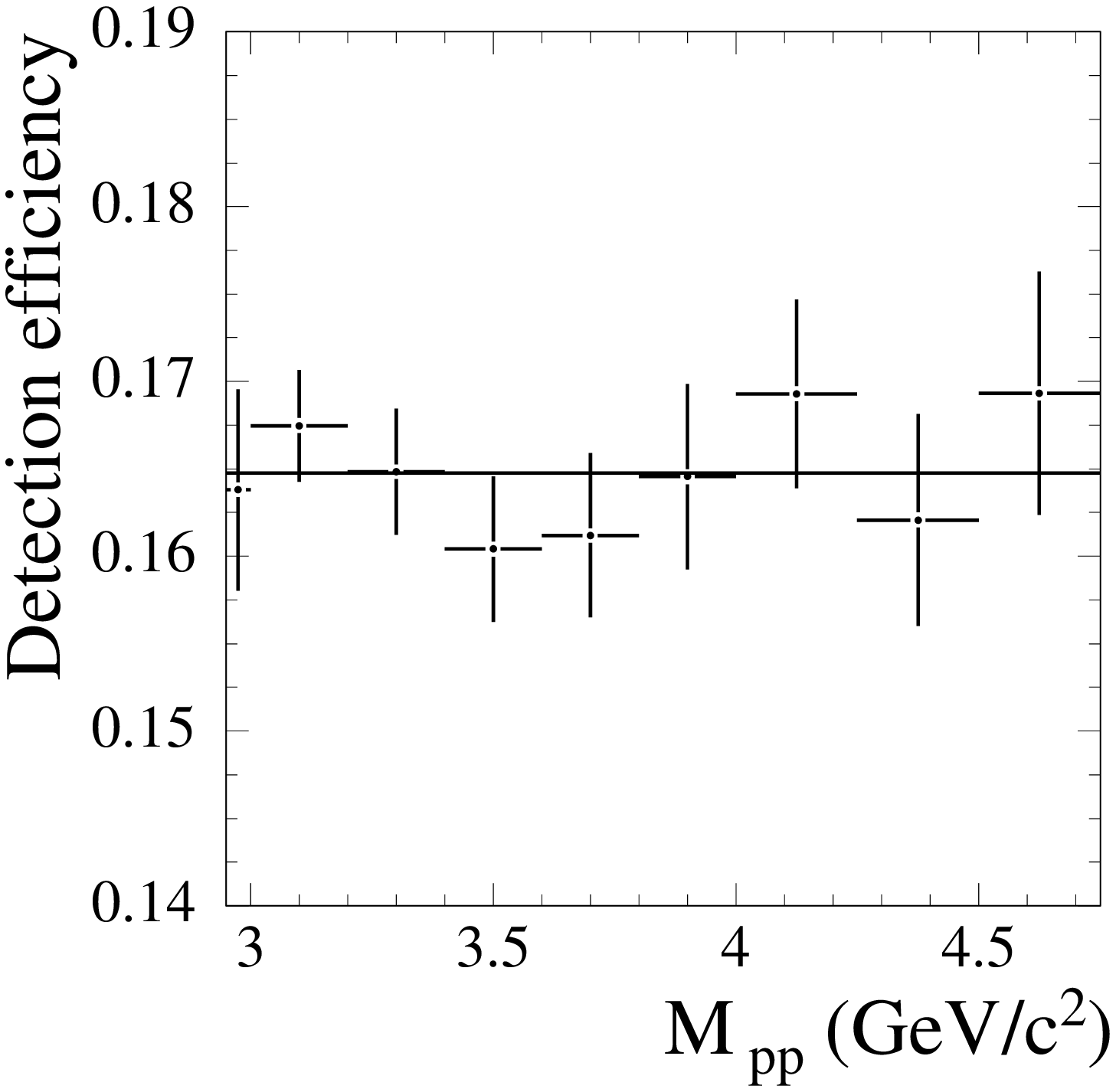}
\caption{The ${p\bar{p}}$ mass dependence of detection
efficiency obtained from MC simulation. The line on the
left plot is the fit to a third-order polynomial. The
efficiency for $M_{p\bar{p}}>3$~GeV/$c^2$ is fit to a
constant value.
\label{detef}}
\end{figure}

\begin{figure}
\psfrag{pp}{$\mbox{p}\bar{\mbox{p}}$}
\includegraphics[width=.48\textwidth]{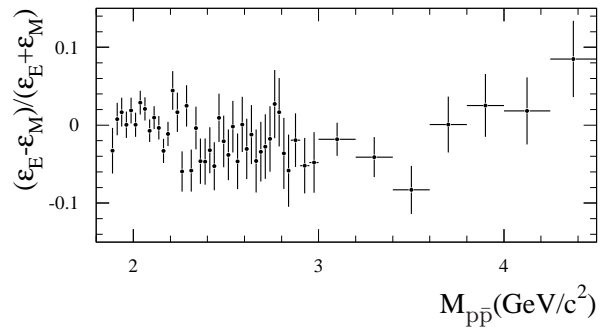}
\caption{The relative difference between
detection efficiencies for a  purely electric ($G_M=0$)
and purely magnetic ($G_E=0$) transition in
$e^+e^-\to p\bar{p} \gamma$ reaction.
\label{efdif}}
\end{figure}

The detection efficiency, determined using Monte Carlo 
simulation, is the ratio of true ${p\bar{p}}$ mass distributions
computed after and before applying selection criteria. Because the
$e^+e^-\to p\bar{p} \gamma$ differential cross-section depends
on  the form factors, the detection efficiency 
is somewhat model-dependent.
The model used in this study has the $|G_E/G_M|$ ratio obtained from a fit of
experimental angular distributions (curve in Fig.~\ref{gegm})
for $M_{p\bar{p}}<3$~GeV/$c^2$,  and $|G_E/G_M|=1$ for higher masses.
The detection efficiency calculated in this model, shown in
Fig.~\ref{detef}, is fit to a third-order polynomial for
$M_{p\bar{p}}<3$~GeV/$c^2$ and a constant for $M_{p\bar{p}}>3$~GeV/$c^2$.
The statistical error of the detection efficiency is about 1\%.
The model error is determined from the uncertainty in the $|G_E/G_M|$ ratio:
for $M_{p\bar{p}}<3$~GeV/$c^2$, varying the ratio within its experimental
uncertainty leads to a 1\% change in the detection efficiency. This
is taken as the model error. This small value is not surprising, due to the
relatively small difference between the detector sensitivities for
pure electric and magnetic transitions. This difference was
calculated with simulated event samples in which $G_E=0$ and
$G_M=0$ and is shown as a function of $M_{p\bar{p}}$ in Fig.~\ref{efdif}.
It does not exceed 20\%.
For masses above 3~GeV/$c^2$, where the $|G_E/G_M|$ ratio
is unknown, a 10\% model error equal to
half of the difference between detection efficiencies
corresponding $G_M=0$ and $G_E=0$ is used.

The efficiency determined from MC simulation
($\varepsilon_{MC}$) must be corrected to account for
data-MC simulation differences in detector response:
\begin{equation}
\varepsilon=\varepsilon_{MC}\prod (1+\delta_i),
\label{eq_eff_cor}
\end{equation}
where $\delta_i$ are efficiency corrections
for each of several effects. These corrections are
discussed in detail below and summarized in Table~\ref{tab_ef_cor}.

Inaccuracy in the simulation of angular and momentum resolutions
and radiative corrections may account for some of the data-MC
difference in the fraction of events rejected by the requirement that
$\chi^2_p<30$.
The efficiency correction for this effect is estimated by comparing
data and simulated $\chi^2$ distributions for the $e^+e^-\to \mu^+\mu^-\gamma$
process, which has kinematics similar to the process under study.
An exclusive $e^+e^-\to \mu^+\mu^-\gamma$ sample is  selected by requiring that
both charged tracks be identified as muons. To remove possible
background contributions from hadronic events with $J/\psi\to \mu^+\mu^-$
 decay,  events with  di-muon invariant mass in the range
$3.0 <M_{\mu\mu}<3.2$~GeV/$c^2$ are excluded.
The ratio of the number of
selected muon events with $\chi^2_\mu>30$ and $\chi^2_\mu<30$
varies from 0.35 to 0.4 in the
$M_{p \bar p}$ range from threshold to 4.5~GeV/$c^2$.
When comparing data and MC simulation in the region  
$\chi^2_\mu>30$,
the cut $M_{\mu\mu\gamma}<8$~GeV/$c^2$ needs to be applied, for
consistency, to the data sample,  since this cut is already applied into the
MC simulation and therefore some events of this non-signal region are
rejected in the simulated sample.
To characterize data-MC simulation difference in the $\chi^2$ distribution,
a double ratio ($\kappa$) is calculated as the ratio 
of $N(\chi^2_\mu>30)/N(\chi^2_\mu<30)$ 
obtained from data to the same quantity obtained from MC simulation.
The value of the double ratio  is 
$\kappa = 1.04\pm0.01$, essentially independent of mass.
The efficiency correction for the $\chi^2$ cut is calculated as
\begin{equation}
\delta_1=\frac{N(\chi^2<30)+N(\chi^2>30)}{N(\chi^2<30)+\kappa
N(\chi^2>30)}-1,  \end{equation}
where $N(\chi^2<30)$ and $N(\chi^2>30)$ are the numbers of simulated
$p\bar{p}\gamma$ events with
$\chi^2<30$ and $\chi^2>30$, respectively.
The values of the efficiency correction $\delta_1$
for different $p\bar{p}$ masses are listed in Table~\ref{tab_ef_cor}.
Its statistical  error is about 0.3\%. An additional
1\% systematic error, equal to
the correction variation in the $p\bar{p}$ mass region of
interest, is added in quadrature.

The effect of the $\chi^2_K>30$ cut is studied using 
$e^+e^-\to J/\psi\gamma \to p\bar{p}\gamma$ events. The number of
$J/\psi$'s is determined using the sideband subtraction method.
The event losses due to $\chi^2_K>30$ cut are found to be
$1.7\pm0.7\%$ in data and $1.7\pm0.2\%$ in MC simulation.
As the data and simulated values are in good agreement, there is no 
need to  introduce any efficiency corrections for the $\chi^2_K>30$ cut.
The systematic uncertainty associated with this cut is 0.7\%. 

      Another possible source of data-MC simulation differences is track loss.
The systematic uncertainty due to differences in track reconstruction 
is estimated to be 1.3\% per track. Specifically, for $e^+e^-\to p\bar{p}\gamma$ 
  only, the systematic error can originate from slightly imperfect simulation
of nuclear interactions of protons and antiprotons in 
the material before the SVT and DCH. The simulation shows that
nuclear interaction leads to the loss of approximately
6\% of $e^+e^-\to p\bar{p}\gamma$ 
events. For data-MC simulation comparison,  a specially
selected event sample with $\Lambda(\bar{\Lambda})$ decaying
into $p(\bar{p})\pi$ is used. The $\Lambda$ are selected by imposing
requirements on $p\pi$ invariant mass and the $\Lambda$ flight distance.
The amount of material before the SVT
(1.5\% of nuclear interaction length) is comparable
to the amount of material between the SVT and the DCH 
(1.4\% of nuclear interaction length).
The probability of track losses between the SVT and the DCH
is measured from the $\Lambda(\bar{\Lambda})$ sample.
The data and simulation probabilities are found to be in
good agreement for protons. 
A substantial difference is observed
for antiprotons, which is consistent with a
large (a factor of $2.6\pm1.0$) overestimation of the antiproton annihilation 
cross-section in simulation. 
This difference in the antiproton
annihilation cross-section in data and simulation leads to a correction of
about $(1.0\pm0.4)\%$ to the detection efficiency for $p\bar{p}\gamma$ events. 

The data-MC simulation difference in the particle identification
is studied with use of events with a $J/\psi\to p\bar{p}$ decay.
Due to the narrow $J/\psi$ width
and low background, the  number of $J/\psi\to p\bar{p}$ decays
may  be determined using selections with either one or two identified
protons.    The background from non-$J/\psi$ events is
subtracted using sidebands.
The $p / \bar{p}$ identification probabilities
are determined as functions of the $p / \bar{p}$ momenta
by calculating the ratio of the number of events
with both the proton and the antiproton identified to the number of
events with only one identified proton or antiproton.
The ratio of data-MC identification probabilities is used to reweight
selected simulated events and calculate efficiency corrections.
The correction is about $(3\pm 3)\%$ and varies within 0.5\%
depending on $p\bar{p}$ mass. The error in the correction factor is determined 
from the statistical 
uncertainty in number of selected $J/\psi$ events.

Another correction must be applied to the photon detection efficiency. 
There are two main sources for this correction: data-MC simulation 
differences in the probability of photon conversion in the detector
material before the DCH, and the effect of dead calorimeter channels.
A sample of $e^+e^-\to \mu^+\mu^-\gamma$ events is 
used to determine the photon 
inefficiency in data. 
Events with exactly two 
oppositely charged 
tracks identified as muons are selected and a kinematic fit is performed,
constraining zero recoil mass  against the muon pair. A tight cut on $\chi^2$ of the kinematic fit  selects events with only one photon in the final 
state. The photon  direction is determined from the fit.  The photon
detection inefficiency is calculated using the ratio of number of events not
passing the  $E_\gamma^{CM} > 3$~GeV cut
to the total number of selected $\mu^+\mu^-\gamma$ events.
The obtained photon inefficiency
3.3\% can be compared to the 2\% inefficiency in $e^+e^-\to p\bar{p}\gamma$
simulation. The observed data-MC difference
in the photon inefficiency leads to an efficiency
       correction of $(-1.3\pm0.1)\%$
that is practically independent of $p\bar{p}$ mass.
The data-MC simulation difference in the probability of photon conversion
is studied using $e^+e^-\to \gamma\gamma$ events 
and  found to be $(0.4\pm0.2)\%$.

The quality of the simulation of the trigger efficiency is also studied. The
overlap of the samples of events passing different
trigger criteria and the independence of these triggers are used to measure
the trigger efficiency. A small difference ($(-0.6\pm0.3)\%$)
in trigger efficiency between data and MC simulation is observed
for $p\bar{p}$ masses below 2.025~GeV/$c^2$.

All efficiency corrections are summarized in Table~\ref{tab_ef_cor}.
The corrected detection efficiencies are listed in
Table~\ref{sumtab}. The uncertainty in  detection efficiency
includes a simulation statistical error, a model uncertainty,
and the uncertainty of the efficiency correction.
\begin{table}
\caption{The values of different efficiency corrections $\delta_i$
for $p \bar{p}$ masses 1.9, 3.0, and 4.5~GeV/$c^2$.
\label{tab_ef_cor}}
\begin{ruledtabular}
\begin{tabular}{lccc}
 effect         &$\delta_i(1.9),\%$&$\delta_i(3),\%$&$\delta_i(4.5),\%$\\
 $\chi^2_p < 30$ cut   & $-0.7\pm1.0$  &$-1.1\pm1.0$ &$-1.7\pm1.0$\\
 $\chi^2_K > 30$ cut   & $0.0\pm0.7$   &$0.0\pm0.7$  &$0.0\pm0.7$\\
 track reconstruction& $0.0\pm3.0$  & $0.0\pm3.0$ & $0.0\pm3.0$\\
 nuclear interaction & $0.8\pm0.4$  & $1.1\pm0.4$ & $1.0\pm0.4$\\
 PID                 & $2.5\pm3.3$  & $3.2\pm2.4$ & $3.5\pm2.7$\\
 photon inefficiency &$-1.3\pm0.1$  &$-1.3\pm0.1$ &$-1.3\pm0.1$\\
 photon conversion   & $0.4\pm0.2$  & $0.4\pm0.2$ & $0.4\pm0.2$\\
 trigger             &$-0.6\pm0.3$  &     --      &    --      \\
 \hline
 total               & $1.1\pm4.7$  & $2.3\pm4.1$ & $1.9\pm4.2$\\
  \end{tabular}
\end{ruledtabular}
 \end{table}
\section{\boldmath $e^+e^-\to p\bar{p}$
cross-section and proton form factor}
The cross-section for $e^+e^-\to p\bar{p}$ is calculated from
the $p\bar{p}$ mass spectrum using expression
\begin{equation}
\sigma_{p\bar{p}}(m)=\frac{({\rm d}N/{\rm d}m)_{corr}}{\varepsilon\, R\,
{\rm d}L/{\rm d}m},
\end{equation}
where $m$ is the $p\bar{p}$ invariant mass,
$({\rm d}N/{\rm d}m)_{corr}$ is the mass spectrum corrected for
resolution effects,
${{\rm d}L}/{{\rm d}m}$ is the ISR differential
luminosity, $\varepsilon$ is the detection efficiency as a function of mass,
and $R$ is a radiative correction factor accounting for the Born mass
spectrum distortion due to emission of several photons by the initial
       electron or positron. The ISR luminosity is calculated
using the total integrated luminosity $L$ and the probability density
function for ISR photon emission in Eq.~\ref{eq2}:
\begin{equation}
\frac{{\rm d}L}{{\rm d}m}=\frac{\alpha}{\pi x}\left(
(2-2x+x^2)\log\frac{1+C}{1-C}-x^2 C\right)\frac{2m}{s}\,L.
\label{ISRlum}
\end{equation}
Here $x=1-m^2/s$, $\sqrt{s}$ is the $e^+e^-$ center-of-mass energy,
$C=\cos{\theta_0^\ast}$,
and $\theta_0^\ast$ determines the range of polar angles in the 
$e^+e^-$ center-of-mass frame:
$\theta_0^\ast<\theta_\gamma^\ast<180^\circ-\theta_0^\ast$ 
      for the ISR photon. In this study, $\theta_0^\ast=20^\circ$,
because the detector efficiency is determined using a simulation with
$20^\circ<\theta_\gamma^\ast<160^\circ$. The  integrated ISR luminosity
for each $M_{p\bar{p}}$ bin is listed in Table~\ref{sumtab}.

The radiative correction factor $R$ is determined using Monte Carlo
simulation at the generator level, without any detector simulation.
Two  $p\bar{p}$ mass spectra are generated; the first using
the pure Born amplitude for the process $e^+ e^- \to p\bar{p}\gamma $,
and the second  using a model with higher-order
radiative corrections included with the structure function
method~\cite{strfun}.
The radiative correction factor $R$ is the ratio of the second spectrum
to the first, and
varies from 1.002 at the  $p\bar{p}$ threshold to 1.02 at 4.5~GeV mass.
The value of $R$ depends on the requirement
on the invariant mass of the $p\bar{p}\gamma$ system.
The $R$ in this study corresponds to the requirement
$M_{p\bar{p}\gamma} > 8$~GeV/$c^2$ imposed in  the simulation.
The theoretical uncertainty in the radiative correction calculation
with the structure function method does not exceed 1\%~\cite{strfun}.
To check  the theoretical uncertainty, a comparison of
the cross-sections calculated with the structure function method
and the Phokhara~\cite{kuhn_pp} event generator is performed. The
Phokhara generator
uses formulae with next-to-leading order radiative corrections
in the initial state.  The uncertainty of Phokhara generator is
estimated to be less than 1\%~\cite{phokhara}.
The ratio of $p\bar{p}$ mass spectra obtained with the two generators
differs from unity by about 1\% and does not contradict
estimates of the theoretical uncertainties.
The radiative corrections calculated include  
initial state radiation and the effect
of loops at electron vertex, but do not include
corrections for leptonic and hadronic vacuum polarization
in the photon propagator.
Cross-sections obtained with such corrections are
sometimes referred to as ``dressed'' cross-sections
while those which account fully for higher order
processes are referred to as ``bare'' cross-sections.
See Ref.\cite{DE} for a more complete discussion.

\begin{figure}
\psfrag{pp}{$\mbox{p}\bar{\mbox{p}}$}
\includegraphics[width=.45\textwidth]{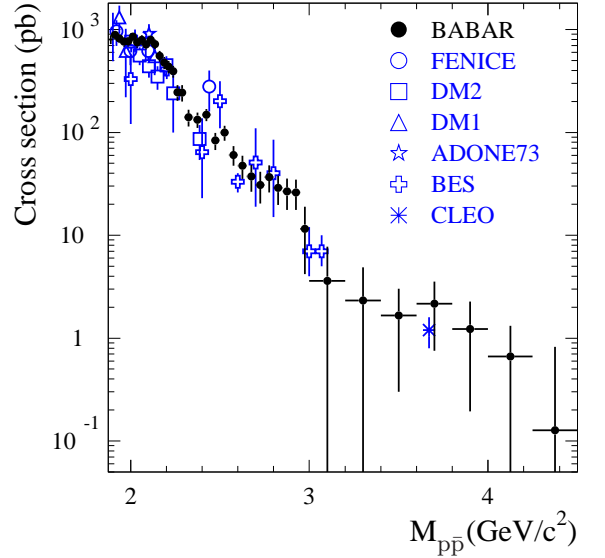}
\caption{The $e^+e^-\to p\bar{p}$ cross-section measured in this work and
$e^+e^-$ experiments: FENICE\cite{FENICE}, DM2\cite{DM2}, DM1\cite{DM1},
ADONE73\cite{ADONE73}, BES\cite{BES}, CLEO\cite{CLEO}.
The contribution of $J/\psi\to p\bar{p}$ and $\psi (2S)\to p\bar{p}$
decays is subtracted.} 
\label{csb}
\end{figure}
\begin{figure}
\psfrag{pp}{$\mbox{p}\bar{\mbox{p}}$}
\includegraphics[width=.45\textwidth]{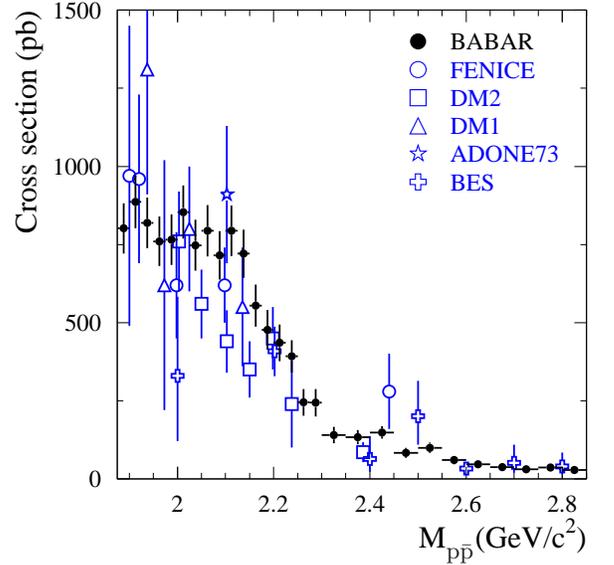}
\caption{The $e^+e^-\to p\bar{p}$ cross-section near threshold 
measured in this work and
$e^+e^-$ experiments: FENICE\cite{FENICE}, DM2\cite{DM2}, DM1\cite{DM1},
ADONE73\cite{ADONE73}, BES\cite{BES}.
\label{cscomp}}
 \end{figure}

 \begin{table*}
\caption{ $p\bar{p}$ invariant mass ($M_{p\bar{p}}$),
number of selected events ($N$) with  background subtracted,
detection efficiency ($\varepsilon$),
ISR luminosity ($L$), measured cross-section
($\sigma_{p\bar{p}}$), and $F_p$, the effective form factor for $e^+e^-\to p\bar{p}$.
The contribution of $J/\psi\to p\bar{p}$ and $\psi (2S)\to p\bar{p}$
decays has been subtracted.
The quoted uncertainties in $N$ and $\sigma$ are statistical and systematic.
For the form factor, the combined uncertainty is listed.
\label{sumtab}} 
\begin{ruledtabular}
\begin{tabular}{cccccc}
 $M_{p\bar{p}}$ (GeV/$c^2$)&$N$&$\varepsilon$&$L$ (pb$^{-1}$)&$\sigma_{p\bar{p}}$ (pb)&$|F_p|$\\
1.877--1.900&$157\pm13\pm3$&$0.171\pm0.008$& 1.141&$802\pm68\pm43$&$0.453_{-0.025}^{+0.023}$\\
1.900--1.925&$190\pm15\pm3$&$0.173\pm0.008$& 1.236&$887\pm71\pm46$&$0.354_{-0.017}^{+0.017}$\\
1.925--1.950&$180\pm15\pm3$&$0.175\pm0.008$& 1.254&$819\pm68\pm43$&$0.305_{-0.015}^{+0.015}$\\
1.950--1.975&$171\pm15\pm4$&$0.177\pm0.008$& 1.272&$760\pm68\pm41$&$0.276_{-0.015}^{+0.014}$\\
1.975--2.000&$176\pm16\pm5$&$0.178\pm0.008$& 1.290&$765\pm68\pm44$&$0.266_{-0.015}^{+0.014}$\\
2.000--2.025&$201\pm17\pm5$&$0.180\pm0.008$& 1.308&$854\pm71\pm48$&$0.273_{-0.014}^{+0.013}$\\
2.025--2.050&$181\pm16\pm6$&$0.182\pm0.008$& 1.328&$748\pm68\pm45$&$0.250_{-0.014}^{+0.013}$\\
2.050--2.075&$196\pm17\pm6$&$0.183\pm0.008$& 1.346&$794\pm68\pm46$&$0.254_{-0.014}^{+0.013}$\\
2.075--2.100&$180\pm17\pm5$&$0.184\pm0.008$& 1.365&$715\pm66\pm40$&$0.239_{-0.013}^{+0.013}$\\
2.100--2.125&$203\pm18\pm5$&$0.185\pm0.008$& 1.383&$794\pm68\pm44$&$0.250_{-0.013}^{+0.012}$\\
2.125--2.150&$188\pm17\pm5$&$0.185\pm0.009$& 1.402&$721\pm66\pm40$&$0.237_{-0.013}^{+0.012}$\\
2.150--2.175&$147\pm15\pm5$&$0.186\pm0.009$& 1.421&$554\pm58\pm34$&$0.207_{-0.013}^{+0.012}$\\
2.175--2.200&$128\pm15\pm6$&$0.187\pm0.009$& 1.440&$477\pm56\pm31$&$0.191_{-0.013}^{+0.012}$\\
2.200--2.225&$119\pm14\pm5$&$0.187\pm0.009$& 1.459&$435\pm52\pm29$&$0.183_{-0.013}^{+0.012}$\\
2.225--2.250&$109\pm13\pm3$&$0.187\pm0.009$& 1.478&$392\pm47\pm21$&$0.174_{-0.012}^{+0.011}$\\
2.250--2.275&$ 69\pm11\pm3$&$0.188\pm0.008$& 1.497&$245\pm40\pm16$&$0.137_{-0.013}^{+0.012}$\\
2.275--2.300&$ 70\pm11\pm5$&$0.188\pm0.008$& 1.516&$244\pm39\pm20$&$0.137_{-0.013}^{+0.012}$\\
2.300--2.350&$ 82\pm12\pm9$&$0.188\pm0.008$& 3.092&$140\pm21\pm16$&$0.105_{-0.010}^{+0.009}$\\
2.350--2.400&$ 80\pm11\pm7$&$0.188\pm0.008$& 3.172&$133\pm19\pm13$&$0.103_{-0.009}^{+0.008}$\\
2.400--2.450&$ 91\pm11\pm2$&$0.187\pm0.008$& 3.251&$149\pm18\pm 7$&$0.110_{-0.008}^{+0.007}$\\
2.450--2.500&$ 52\pm 9\pm2$&$0.187\pm0.008$& 3.331&$ 83\pm15\pm 5$&$0.083_{-0.008}^{+0.008}$\\
2.500--2.550&$ 63\pm10\pm2$&$0.186\pm0.008$& 3.414&$100\pm16\pm 6$&$0.092_{-0.008}^{+0.007}$\\
2.550--2.600&$ 39\pm 8\pm2$&$0.185\pm0.008$& 3.496&$ 60\pm13\pm 4$&$0.072_{-0.008}^{+0.007}$\\
2.600--2.650&$ 31\pm 8\pm2$&$0.183\pm0.008$& 3.580&$ 47\pm11\pm 4$&$0.065_{-0.009}^{+0.008}$\\
2.650--2.700&$ 25\pm 7\pm2$&$0.182\pm0.008$& 3.664&$ 37\pm10\pm 4$&$0.059_{-0.009}^{+0.008}$\\
2.700--2.750&$ 21\pm 7\pm2$&$0.180\pm0.008$& 3.749&$ 31\pm10\pm 4$&$0.054_{-0.010}^{+0.008}$\\
2.750--2.800&$ 25\pm 7\pm2$&$0.179\pm0.008$& 3.837&$ 37\pm10\pm 4$&$0.060_{-0.010}^{+0.008}$\\
2.800--2.850&$ 20\pm 6\pm2$&$0.178\pm0.008$& 3.924&$ 30\pm 9\pm 3$&$0.054_{-0.009}^{+0.008}$\\
2.850--2.900&$ 19\pm 6\pm2$&$0.176\pm0.008$& 4.013&$ 27\pm 9\pm 3$&$0.052_{-0.010}^{+0.008}$\\
2.900--2.950&$ 19\pm 6\pm2$&$0.175\pm0.007$& 4.103&$ 26\pm 8\pm 3$&$0.052_{-0.009}^{+0.008}$\\
2.950--3.000&$  9\pm 5\pm2$&$0.173\pm0.007$& 4.195&$ 12\pm 7\pm 3$&$0.035_{-0.014}^{+0.010}$\\
3.000--3.200&$ 11\pm 9\pm8$&$0.169\pm0.018$&17.719&$  3.6\pm 3.0\pm 2.8$&$0.021_{-0.021}^{+0.009}$\\
3.200--3.400&$  8\pm 5\pm7$&$0.169\pm0.018$&19.289&$  2.3\pm 1.6\pm 2.0$&$0.017_{-0.017}^{+0.008}$\\
3.400--3.600&$  6\pm 4\pm3$&$0.169\pm0.018$&20.960&$  1.7\pm 1.0\pm 0.9$&$0.016_{-0.009}^{+0.005}$\\
3.600--3.800&$  8\pm 4\pm3$&$0.168\pm0.018$&22.739&$  2.2\pm 1.1\pm 0.8$&$0.019_{-0.008}^{+0.005}$\\
3.800--4.000&$  5\pm 3\pm3$&$0.168\pm0.018$&24.645&$  1.2\pm 0.8\pm 0.7$&$0.015_{-0.009}^{+0.005}$\\
4.000--4.250&$  4\pm 3\pm3$&$0.168\pm0.018$&33.701&$  0.7\pm 0.5\pm 0.5$&$0.011_{-0.010}^{+0.005}$\\
4.250--4.500&$  1\pm 3\pm3$&$0.167\pm0.018$&37.214&$  0.1\pm 0.4\pm 0.5$&$0.005_{-0.005}^{+0.008}$\\
\end{tabular}
\end{ruledtabular}
\end{table*}

The resolution-corrected mass spectrum is obtained by unfolding
the mass resolution from the measured mass spectrum. 
Using the MC simulation, the migration matrix $A$ is obtained, 
representing the probability that an event with  true
mass ($M_{p\bar{p}}^{true}$) in bin $j$ is reconstructed in bin $i$ :
\begin{equation}
\left( \frac{{\rm d}N}{{\rm d}m}\right)^{rec}_i=
\sum_j A_{ij}\left(\frac{{\rm d}N}{{\rm d}m}\right)^{true}_j.
\end{equation}
As the chosen bin width significantly exceeds the
mass resolution for all $p\bar{p}$ masses, the
migration matrix is nearly diagonal, with the values of
diagonal elements  $ \sim 0.9$,
and next-to-diagonal $ \sim 0.05$.
We unfold the mass spectrum by applying
the inverse of the migration matrix to the measured
spectrum. This procedure changes 
the shape of the mass distribution insignificantly,
but increases the errors (by $\approx $20\%)
and their correlations. 

        The number of events in each mass bin is listed in Table~\ref{sumtab}.
The quoted errors are statistical and systematic 
(with the systematic errors due to uncertainties in background subtraction).
The calculated cross-section for $e^+e^-\to p\bar{p}$ is
shown in Fig.~\ref{csb}  and listed in Table~\ref{sumtab}.
For mass bins 3--3.2~GeV/$c^2$ and 3.6--3.8~GeV/$c^2$, the 
nonresonant cross-section is quoted with $J/\psi$ and $\psi(2S)$ contributions 
excluded   (see Sec.~\ref{jpsi}).
The errors quoted are statistical and
systematic. The systematic uncertainty includes the uncertainty in 
the number of signal
events and detection efficiency, an error of total integrated luminosity
(1\%), and the  uncertainty in the radiative corrections (1\%).
A comparison of this result with available $e^+e^-$ data is shown
in Fig.~\ref{csb} and the near-threshold region is shown in Fig.~\ref{cscomp}.

\begin{figure}
\psfrag{pp}[][][0.7]{$\mbox{p}\bar{\mbox{p}}$}
\includegraphics[width=.96\linewidth]{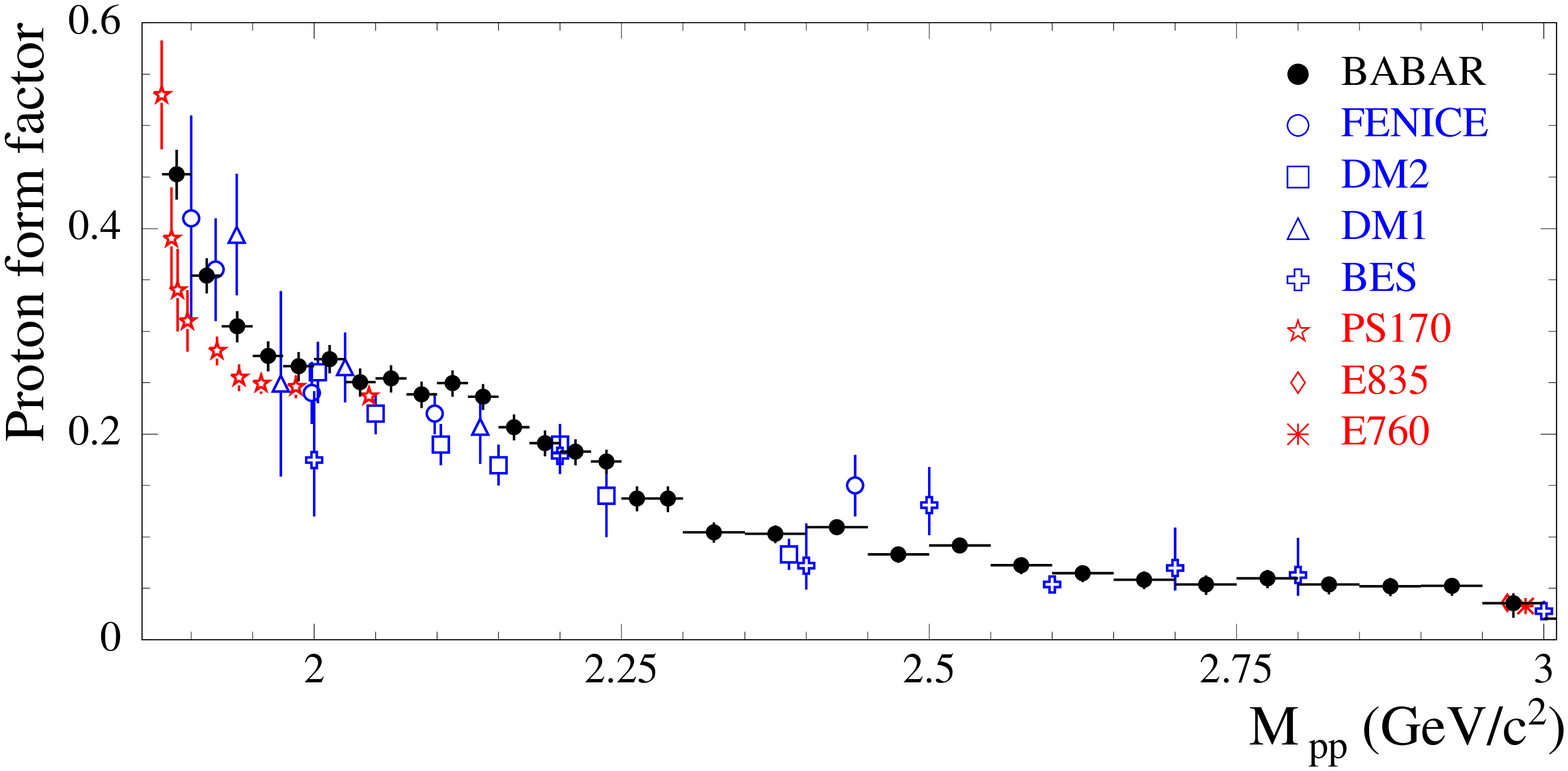}
\includegraphics[width=.96\linewidth]{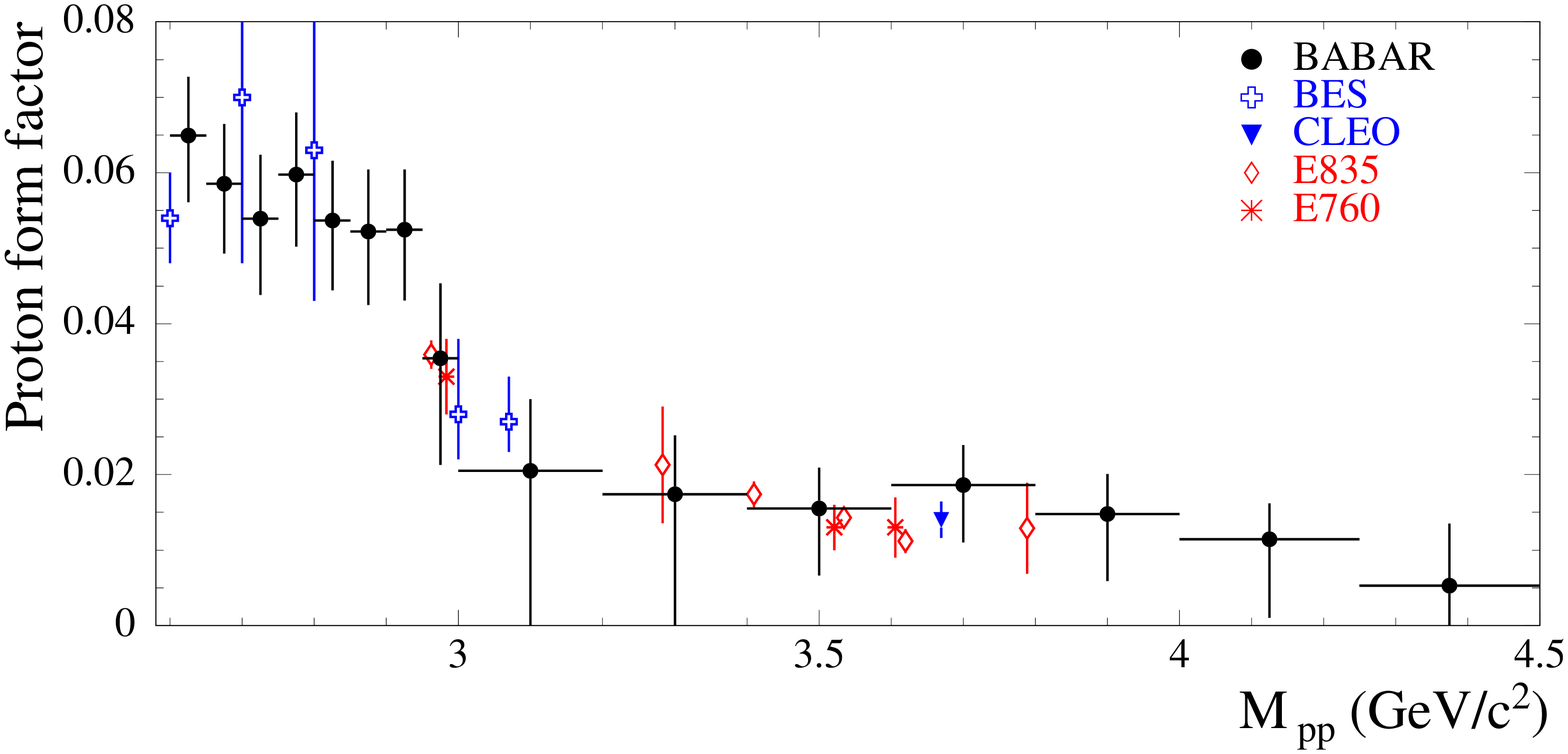}
\caption{The proton effective form factor measured in this work and
in $e^+e^-$ and $p\bar{p}$ experiments: FENICE\cite{FENICE}, DM2\cite{DM2},
DM1\cite{DM1}, BES\cite{BES}, CLEO\cite{CLEO}, PS170\cite{LEAR},
E835\cite{E835}, E760\cite{E760}. The upper plot shows the mass interval
from $p\bar{p}$ threshold to 3.01 GeV/$c^2$. The lower plot presents
data for $p\bar{p}$ masses from 2.58 to 4.50 GeV/$c^2$.
\label{ppffx}}
\end{figure}

\begin{figure}
\psfrag{pp}{$\mbox{p}\bar{\mbox{p}}$}
\includegraphics[width=.48\textwidth]{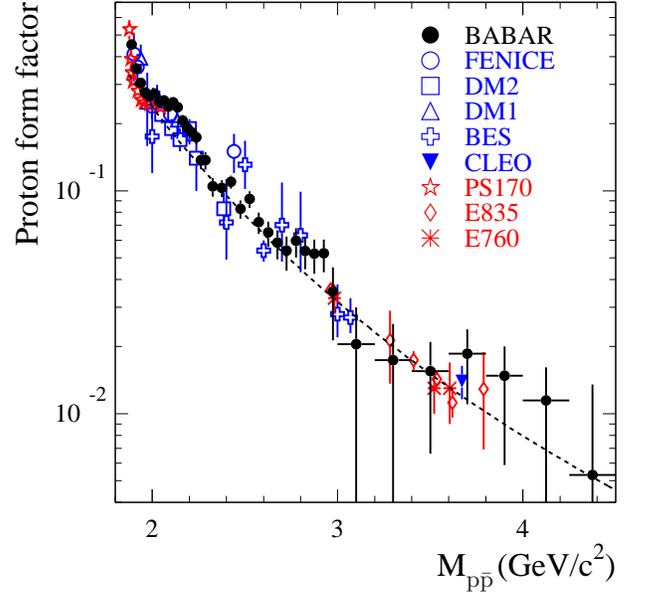}
\caption{The proton effective form factor measured in this work and
in $e^+e^-$ and $p\bar{p}$ experiments, shown on a logarithmic scale: 
FENICE\cite{FENICE}, DM2\cite{DM2}, DM1\cite{DM1}, BES\cite{BES}, CLEO\cite{CLEO},
PS170\cite{LEAR}, E835\cite{E835}, E760\cite{E760}.
The curve corresponds to the QCD fit described in the text.
\label{ppff}}
\end{figure}

\begin{figure}
\psfrag{pp}{$\mbox{p}\bar{\mbox{p}}$}
\includegraphics[width=.48\textwidth]{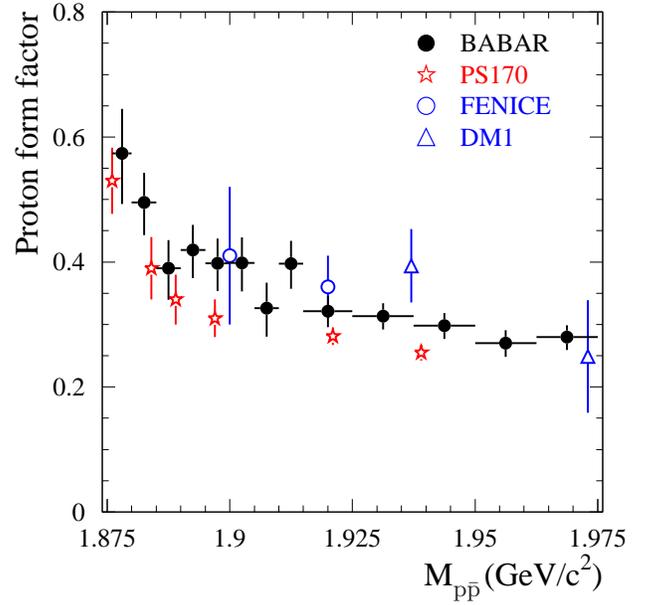}
\caption{The proton effective form factor near $p\bar{p}$ threshold
measured in this work and
in $e^+e^-$ and $p\bar{p}$ experiments: FENICE\cite{FENICE},
DM1\cite{DM1}, PS170\cite{LEAR}.
\label{fflow}}
\end{figure}

The $e^+e^-\to p\bar{p}$ cross-section is a function of two form factors,
but due to poor determination of the  $|G_E/G_M|$ ratio, they cannot 
be extracted from the data simultaneously with reasonable accuracy.
Therefore, the effective form factor is introduced:
\begin{equation}
         |F_p(m)|=\sqrt{\sigma_{p\bar{p}}(m)/\sigma_n(m)},
\end{equation}
where $\sigma_{p\bar{p}}(m)$ is the measured 
$e^+e^-\to p\bar{p}$ cross-section
and $\sigma_n(m)$ is the cross-section
obtained from Eq.~\ref{eq4} under the assumption that $|G_E|=|G_M|=1$.
At $M_{p\bar{p}}=2$~GeV/$c^2$ $\sigma_n \simeq 10~{\rm nb}$. 
This  definition 
of the effective form factor $F_p(m)$ permits comparison of our
measurements  with measurements from
other experiments, in $e^+e^-$ as well as $p\bar{p}$
collisions. 
Most available form-factor data are analyzed using the 
assumption that $|G_E|=|G_M|$. 
The calculated effective form factor is shown in Fig.~\ref{ppffx} 
(linear scale), in  Fig.~\ref{ppff} (logarithmic scale)
and in Table~\ref{sumtab}.
\begin{table*}
\caption{ $p\bar{p}$ invariant mass ($M_{p\bar{p}}$),
number of selected events ($N$) after background subtraction,
measured cross-section
($\sigma_{p\bar{p}}$), and effective form factor for $e^+e^-\to p\bar{p}$.
The quoted errors in $N$ and $\sigma_{p\bar{p}}$ are statistical and systematic.
For the effective form factor, the total combined error is listed.
\label{tablow}}
\begin{ruledtabular}
\begin{tabular}{cccc}
$M_{p\bar{p}}$ (GeV/$c^2$)&$N$&$\sigma_{p\bar{p}}$ (pb) & $|F_p|$ \\
1.8760--1.8800&$ 18\pm 5\pm1$&$ 656\pm161\pm40$&$0.574^{+0.071}_{-0.081}$\\
1.8800--1.8850&$ 34\pm 6\pm1$&$ 808\pm155\pm43$&$0.495^{+0.047}_{-0.052}$\\
1.8850--1.8900&$ 27\pm 6\pm1$&$ 656\pm154\pm36$&$0.390^{+0.045}_{-0.050}$\\
1.8900--1.8950&$ 37\pm 7\pm1$&$ 889\pm174\pm48$&$0.419^{+0.041}_{-0.045}$\\
1.8950--1.9000&$ 38\pm 8\pm1$&$ 901\pm182\pm48$&$0.398^{+0.040}_{-0.044}$\\
1.9000--1.9050&$ 42\pm 9\pm1$&$ 995\pm207\pm56$&$0.399^{+0.041}_{-0.046}$\\
1.9050--1.9100&$ 31\pm 8\pm1$&$ 726\pm186\pm41$&$0.326^{+0.040}_{-0.046}$\\
1.9100--1.9150&$ 49\pm 9\pm1$&$1138\pm210\pm60$&$0.397^{+0.036}_{-0.040}$\\
1.9150--1.9250&$ 69\pm10\pm1$&$ 798\pm116\pm43$&$0.321^{+0.024}_{-0.026}$\\
1.9250--1.9375&$ 91\pm11\pm2$&$ 831\pm102\pm44$&$0.313^{+0.020}_{-0.022}$\\
1.9375--1.9500&$ 90\pm11\pm2$&$ 817\pm104\pm43$&$0.298^{+0.020}_{-0.021}$\\
1.9500--1.9625&$ 80\pm12\pm3$&$ 712\pm105\pm42$&$0.270^{+0.021}_{-0.022}$\\
1.9625--1.9750&$ 91\pm12\pm2$&$ 802\pm105\pm43$&$0.280^{+0.019}_{-0.020}$\\
\end{tabular}
\end{ruledtabular}
\end{table*}
The form factors here are averaged
over bin width, and the four points of PS170~\cite{LEAR} with lowest mass
are all situated within the first bin of the \babar\ measurement.
 For the mass region near threshold
where the form factor changes rapidly with mass, the cross-section
and effective form factor with a smaller bin size are calculated. These
results are   listed in Table~\ref{tablow}. The effective form 
factor is shown in   Fig.~\ref{fflow}.
In Figs.~\ref{ppffx}, ~\ref{ppff} and ~\ref{fflow},
it is evident that the \babar\ effective form factor results are in 
reasonable agreement with those of other experiments. 
The form factor has a complex mass dependence.
The significant increase in form factor as the $p\bar{p}$ threshold 
is approached may be a manifestation
of a $p\bar{p}$ subthreshold resonance~\cite{ppres}.
The rapid
decreases of the form factor and cross-section near 2.25~GeV/$c^2$ and 3~GeV/$c^2$
have not been discussed in the literature.
The dashed line in Fig.~\ref{ppff} corresponds to the asymptotic
QCD fit \cite{QCD} for proton form factor 
$F_{p\bar{p}}\sim\alpha_s^2(m^2)/m^4\sim C/(m^4\log^2 (m^2/\Lambda^2))$,
applied to all existing data  with 
$M_{p\bar{p}}>3~{\rm GeV}/c^2$. Here $\Lambda=0.3$~GeV 
and $C$ is free fit parameter. 
It is seen that the asymptotic regime is reached at masses above 3~GeV/$c^2$.
\section{\boldmath The $J/\psi$ and $\psi (2S)$  decays into 
$p\bar{p}$ }\label{jpsi}
The  differential cross-section for ISR production
of a narrow resonance (vector meson $V$),
such as $J/\psi$, decaying into the final state $f$ can be calculated
using~\cite{ivanch}
\begin{equation}
\frac{{\rm d}\sigma(s,\theta_\gamma^\ast)}{{\rm d}\cos{\theta_\gamma^\ast}}
=\frac{12\pi^2 \Gamma(V\to e^+e^-) {\cal B}(V\to f)}{m_V\, s}\,
W(s,x_V,\theta_\gamma^\ast),
\label{eqpsi}
\end{equation}
where $m_V$ and $\Gamma(V\to e^+e^-)$ are the mass and electronic
width of the vector meson $V$, $x_V = 1-{m_V^2}/{s}$,
and ${\cal B}(V\to f)$ is the branching fraction of $V$
into the final state $f$. Therefore, the measurement of the number of
$J/\psi \to p\bar{p}$ decays
in $e^+ e^- \to p\bar{p}\gamma$ determines the product of the
electronic width and the branching fraction:
$\Gamma(J/\psi \to e^+e^-){\cal B}(J/\psi \to p\bar{p})$.
\begin{figure}
\psfrag{pp}[][][0.7]{$\mbox{p}\bar{\mbox{p}}$}
\includegraphics[width=.48\linewidth]{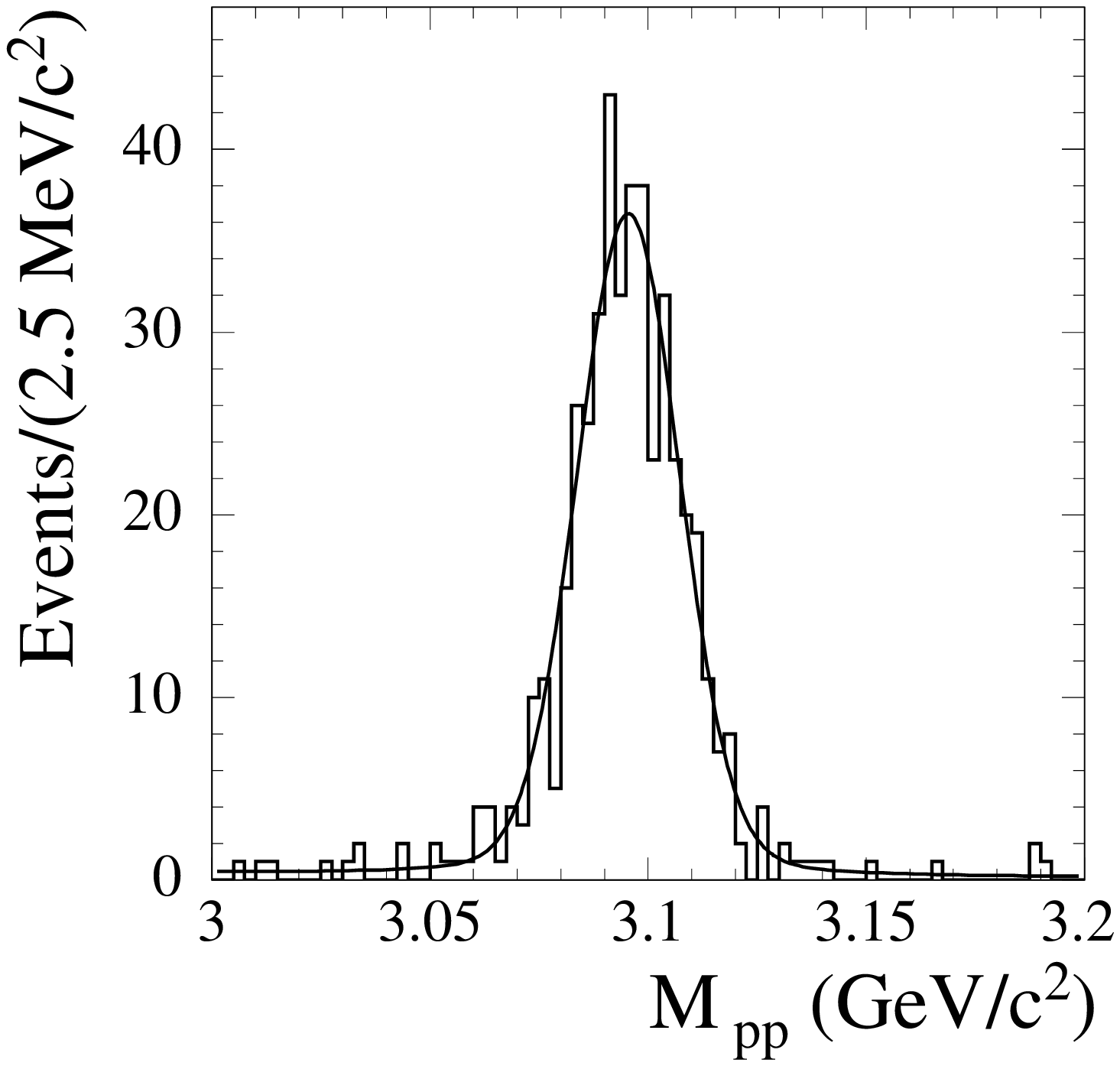}
\includegraphics[width=.48\linewidth]{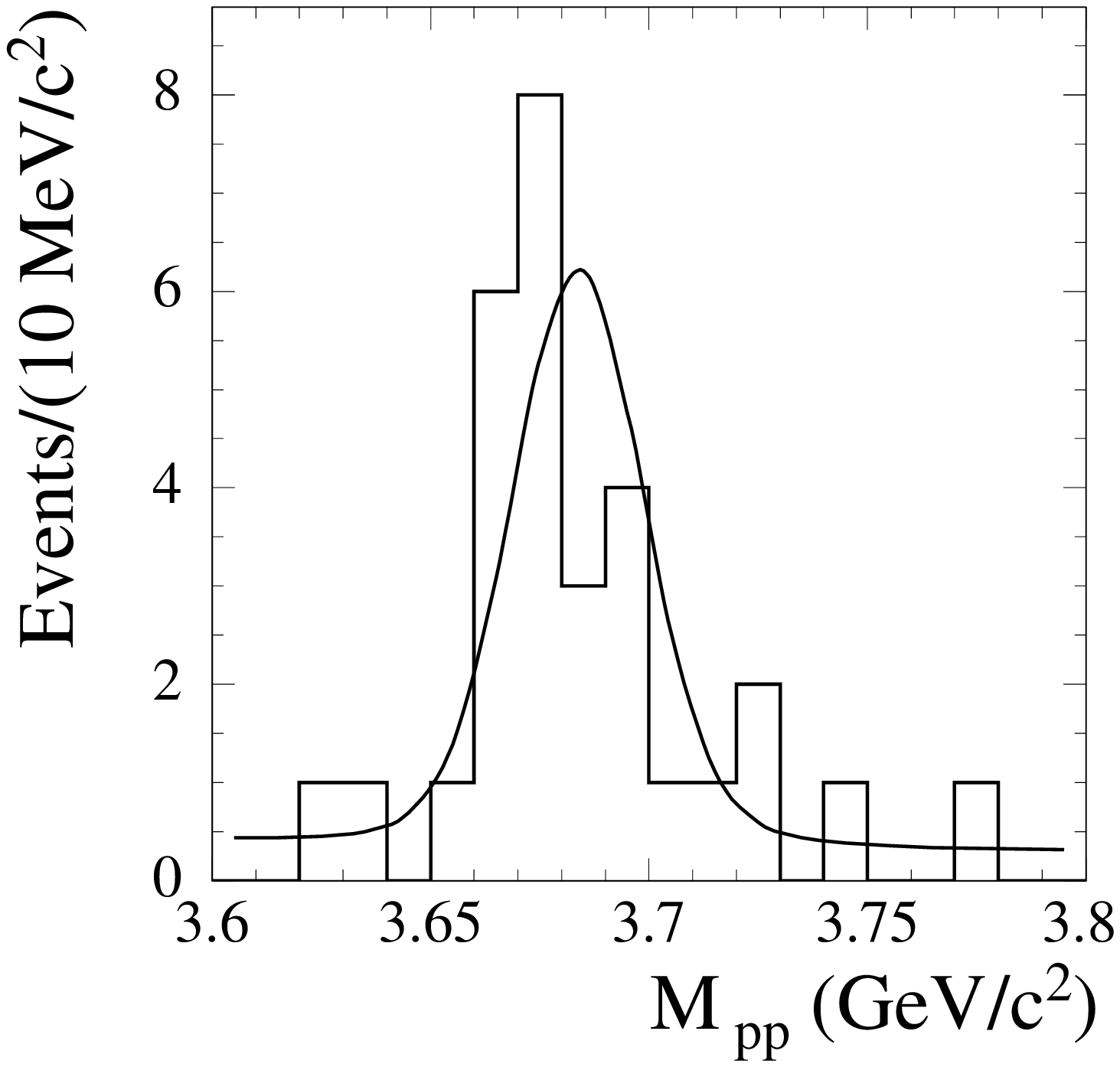}
\caption{The $p\bar{p}$ mass spectra in the mass regions
near $J/\psi$ (left) and $\psi(2S)$ (right). The curves are the
the result of the fit described in the text.
\label{psiexp}}
\end{figure}
The $p\bar{p}$ mass spectra for selected events in the $J/\psi$ and
$\psi(2S)$ mass regions are shown in Fig.~\ref{psiexp}.
To determine the number of resonance events, both spectra
are fitted with a sum of the probability density function (PDF)
for signal plus a linear background.
The resonance PDF is
a  Breit-Wigner function convolved with
a double-Gaussian function describing the detector resolution.
The Breit-Wigner widths and masses for $J/\psi$ and
$\psi(2S)$ are fixed at the world-average values.
The parameters of the resolution function
are determined from simulation. 
To account for possible differences in detector response
between data and simulation,
the simulated resolution function is modified by
adding in quadrature  an additional $\sigma_G$ to both
$ \sigma$'s of the double-Gaussian function and introducing
a shift of the central value of the resonance mass. 
The free parameters in the
fit of $J/\psi$ mass region are the number of resonance events,
the total number of nonresonant background events, 
the slope of background, $\sigma_G$, and mass shift. 
In the $\psi(2S)$ fit the $\sigma_G$ and mass shift values
are fixed at those obtained for the  $J/\psi$.

The fit results are shown as curves in Fig.~\ref{psiexp}. 
Numerically, we find:
$N_{J/\psi}=438\pm22$ and $N_{\psi(2S)}=22.2\pm5.7$;
the number of nonresonant events is $27\pm8$ for the 3--3.2~GeV/$c^2$
mass interval and $7.9\pm4.0$ for the 3.6--3.8~GeV/$c^2$ interval.
These values are used to extract the nonresonant $e^+e^- \to p\bar{p}$
cross-section. 
Since the background subtraction procedure for nonresonant
events (see Sec.~\ref{bkgsub}) uses events with $30<\chi^2_p<60$, 
the mass spectra obtained with this cut may also be fit. The
numbers of $J/\psi$ and nonresonant events are found to be
$27\pm6$ and $6\pm4$. The ratio of $J/\psi$ events with
$30<\chi^2_p<60$ to the number with
$\chi^2_p<30$, $0.061\pm0.014$ is in good
agreement with value of $\beta_{p\bar{p}\gamma}=0.048\pm0.003$ obtained
in Sec.~\ref{bkgsub}. In the $\psi(2S)$ mass region, no events are
selected with $30<\chi^2_p<60$.  The remaining fit parameters are
$\sigma_G=4.2\pm1.8$ MeV/$c^2$ and
$M_{J/\psi}-M_{J/\psi}^{MC}=-(1.8\pm0.7)$~GeV/$c^2$.
The fitted value of $\sigma_G$ leads to a change in simulation
resolution (11 MeV/$c^2$) of 8\%.

The detection efficiency is estimated from MC simulation.
The event generator uses  experimental data for the angular
distribution of protons in $\psi \to p\bar{p}$ decays.
This distribution is described by $1+\alpha \cos^2\vartheta$
with $\alpha=0.660\pm0.045$ for $J/\psi$~\cite{psipp1,BESpp}
and $0.67\pm 0.15$ for $\psi(2S)$~\cite{E835psi}. The model
error in the detection efficiency due to the uncertainty of
$\alpha$ is negligible. The efficiencies are found to be
$\varepsilon_{MC}$ = 0.168$\pm$0.002 for $J/\psi$ and
$\varepsilon_{MC}$ = 0.161$\pm$0.003 for $\psi(2S)$.
The data-MC simulation differences
discussed earlier are used to correct
the former efficiency values by $(2.3\pm 4.0)\%$.

The cross-section for
$e^+e^-\to \psi\gamma\to p\bar{p}\gamma$ for
$20^\circ<\theta_\gamma^\ast<160^\circ$
is calculated as
$$\sigma(20^\circ<\theta_\gamma^\ast<160^\circ)=\frac{N_{\psi}}
{\varepsilon\, R\, L},$$
yielding $(11.0\pm0.6\pm0.5)$ fb and $(0.57\pm0.14\pm0.03)$ fb
for $J/\psi$ and $\psi(2S)$, respectively.
The radiative-correction factor $R=\sigma/\sigma_{Born}$, is
$1.007\pm0.010$ for $J/\psi$ and $1.011\pm0.010$ for $\psi(2S)$,
obtained from a MC simulation at the generator level.

The total integrated luminosity for the data sample  is
$(232\pm 3)$ fb$^{-1}$.
From the measured cross-sections and Eq.~\ref{eqpsi},
the following products are determined:
$$\Gamma(J/\psi\to e^+e^-){\cal B}(J/\psi\to p\bar{p})=
(12.0\pm 0.6\pm 0.5)\mbox{ eV},$$
$$\Gamma(\psi(2S)\to e^+e^-){\cal B}(\psi(2S)\to p\bar{p})=
(0.70\pm0.17\pm0.03)\mbox{ eV}.$$
The systematic errors include the uncertainties of the detection efficiencies,
the integrated luminosity, and the radiative corrections.

Using the world-average values for the electronic widths~\cite{pdg},
we calculate the $\psi\to p\bar{p}$ branching fractions to be
$${\cal B}(J/\psi\to p\bar{p})=(2.22\pm0.16)\times 10^{-3}$$
and
$${\cal B}(\psi(2S)\to p\bar{p})=(3.3\pm0.9)\times 10^{-4}.$$
These values are in agreement with the corresponding world-average values:
$(2.17\pm0.08)\times 10^{-3}$~\cite{pdg} and
$(2.67\pm0.15)\times 10^{-4}$~\cite{pdg,E760psi,CLEOpsi}.

\section{Upper limit on $Y(4260)\to p\bar{p}$ decay}
Recently, a resonant-like structure in the invariant mass spectrum of
$J/\psi \pi^+\pi^-$ near 4.26 GeV/$c^2$ was observed by \babar\ in the
ISR process $e^+e^- \to J/\psi \pi^+\pi^- \gamma$~\cite{yexp}. This structure
can be characterized by a single resonance with a width of about 90~MeV and
is referred to as $Y(4260)$. 
From the $J/\psi \pi^+\pi^-$ mass spectrum,
the $e^+e^- \to Y(4260) \to J/\psi \pi^+\pi^-$ cross section at the maximum
of the $Y(4260)$ resonance was found to be $(51\pm12)$~pb. 
From the fact that the $Y(4260)$ resonance is not observed 
in the total $e^+e^-\to hadrons$ cross section, one can
conclude that $\Gamma(Y(4260)\to e^+e^-)$ is much smaller 
than the corresponding partial widths of
all known $J^{PC}=1^{- -}$ charmonium resonances,
 while $\Gamma(Y(4260)\to J/\psi \pi^+\pi^-)$ is
much larger~\cite{yhybrid}. The four-quark~\cite{y4quark},
hybrid~\cite{yhybrid0,yhybrid}, meson- or baryon-molecular~\cite{ymol1,ymol2}
interpretations
have been suggested to explain these unusual properties of the $Y(4260)$. 
Information about $Y(4260)$ decay modes other than $J/\psi \pi^+\pi^-$
can help clarify the nature of the $Y(4260)$ resonance. In particular,
charmless decays of the $Y(4260)$ are expected in the hybrid model~\cite{yhybrid}.
\begin{figure}
\psfrag{pp}{$\mbox{p}\bar{\mbox{p}}$}
\includegraphics[width=.45\textwidth]{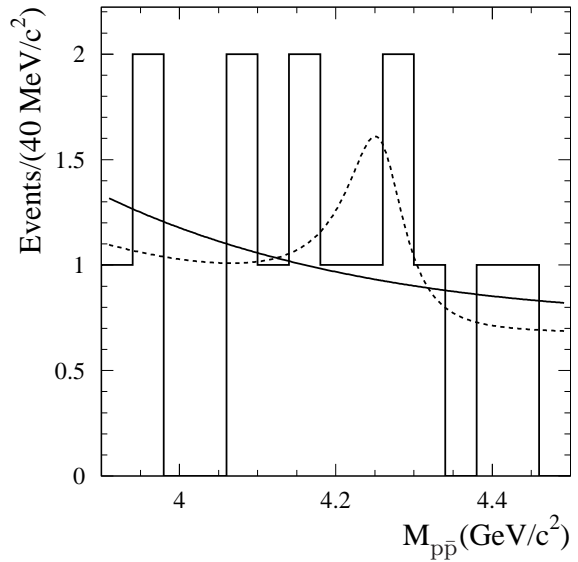}
\caption{The $p\bar{p}$ invariant mass spectrum for
$p\bar{p}\gamma$ candidates in the (3.9--4.5)~GeV/$c^2$ mass range.
The curves are the result of the fit described in the
text.
\label{y4260}}
\end{figure}

The $p\bar{p}$ mass spectrum is shown in Fig.~\ref{y4260}
for $p\bar{p}\gamma$ candidates with $p \bar{p}$ mass in the range
(3.9--4.5)~GeV/$c^2$. 
The mass spectrum is fit with the function
\begin{eqnarray}
\frac{{d}N}{{d}m}=
\frac{{d}L}{{d}m} \varepsilon R
\left |\sqrt{\sigma_0(m)}+
\sqrt{\sigma_Y}\frac{m_Y \Gamma_Y}{m^2-m_Y^2+im_Y \Gamma_Y}\ e^{i\phi} \right |^2 \nonumber \\
\mbox{}+B(m),\phantom{\sqrt{\sigma_0(m)}+ \sqrt{\sigma_Y}\frac{m_Y \Gamma_Y}{m^2-m_Y^2+im_Y \Gamma_Y}\ e^{i\phi}}
\end{eqnarray}
where $\sigma_0(m)$ is the nonresonant cross section for $e^+e^-\to p\bar{p}$, $B$ is
the contribution of background processes,
$\sigma_Y$ is the cross section at the maximum of the Y(4260) resonance,
and $m_Y$ and $\Gamma_Y$ are the resonance
mass and full width, respectively. 
The nonresonant cross section is described by Eq.~\ref{eq4}
with the proton effective form factor parametrized by the PQCD formula
$F_{p\bar{p}}=C/m^4\log^2 (m^2/\Lambda^2)$ with $\Lambda=0.3$~GeV.
The total background contribution in the (3.9--4.5)~GeV/$c^2$ mass range,
mainly from $e^+e^-\to p\bar{p}\pi^0$,
is estimated to be $7.5\pm4.3$ events. 
Due to the large uncertainty in the background, we
cannot determine its slope from the fit;
we assume a uniform background mass distribution. The mass and width
of the Y(4260) are fixed at the values obtained in Ref.~\cite{yexp}: 
$m_Y= (4259\pm10)$~GeV/$c^2$ and
$\Gamma_Y= (88\pm24)$~GeV/$c^2$. 
The values of $C$, $\sigma_Y$, and the interference phase $\phi$ are free in the fit.
The best-fit function is shown in Fig.~\ref{y4260} as the dashed line.
The solid line represents the fit with $\sigma_Y=0$ (null hypothesis). The optimal value of
$\sigma_Y$ is 2.6~pb with a significance of $0.7\sigma$. The significance is
estimated from the ratio of the values
of the likelihood function for the optimal fit and the fit to the null hypothesis. 
Since the best-fit value
of $\sigma_Y$ is compatible with zero, we set an upper limit on the
$e^+e^- \to Y(4260) \to p\bar{p}$ cross section. 

The mass spectrum is fit with different fixed values of the interference phase $\phi$ and
the upper limit at 90\% confidence level (CL)  
is determined as a function of phase with the 
Neyman approach~\cite{Neyman} using a Monte-Carlo technique. 
The upper limit varies from 1.0~pb to 6.4~pb. 
The maximum value, corresponding to $\phi=-\pi/2$, is chosen as a final
result: $\sigma_Y < 6.4$~pb at 90\% CL. 

From the ratio of measured cross sections
for $Y(4260) \to p \bar{p}$ and $Y(4260)\to J/\psi \pi^+\pi^-$, 
we calculate an upper limit on the ratio of branching fractions:
\begin{equation}
\frac{{\cal B}(Y(4260)\to p\bar{p})}{{\cal B}(Y(4260)\to J/\psi\pi^+\pi^-)} < 13\%
\mbox{ at 90\% CL}.
\end{equation}

\section{Summary}
The process $e^+e^-\to p\bar{p}\gamma$ is studied for
$p\bar{p}$ invariant masses up to 4.5~GeV/$c^2$.
From the measured $p\bar{p}$ mass spectrum we extract the
$e^+e^-\to p\bar{p}$ cross-section and proton 
effective form factor.
The form factor has a  complex mass dependence.
  The near-threshold enhancement of the form factor
observed in the PS170 experiment~\cite{LEAR} is confirmed in this study.
  There are also two mass regions, near 2.25~GeV/$c^2$ and 3~GeV/$c^2$,
 that exhibit  steep decreases in the form factor and cross-section.
By analysing the proton angular distributions 
for $M_{p\bar{p}}$ between
threshold and 3~GeV/$c^2$, the ratio $|G_E/G_M|$
is extracted.  
For masses up to 2.1~GeV/$c^2$, this ratio is found to be
significantly greater
than unity, in disagreement with the PS170 measurement~\cite{LEAR}.

From the measured numbers of 
$e^+e^-\to J/\psi\gamma\to p\bar{p}\gamma$ and 
$e^+e^-\to\psi(2S)\gamma \to p\bar{p}\gamma$
events,  the products
$$\Gamma(J/\psi\to e^+e^-){\cal B}(J/\psi\to p\bar{p})=
(12.0\pm 0.6\pm 0.5)\mbox{ eV},$$
$$\Gamma(\psi(2S)\to e^+e^-){\cal B}(\psi(2S)\to p\bar{p})=
(0.70\pm0.17\pm0.03)\mbox{ eV},$$
and their corresponding  branching fractions are determined:
$${\cal B}(J/\psi\to p\bar{p})=(2.22\pm0.16)\times 10^{-3},$$
$${\cal B}(\psi(2S)\to p\bar{p})=(3.3\pm0.9)\times 10^{-4}.$$

The upper limit on $Y(4260) \to p\bar{p}$ decay is obtained
at 90\% CL:
$$\frac{{\cal B}(Y\to p\bar{p})}{{\cal B}(Y\to J/\psi\pi^+\pi^-)} < 13\%.$$

\section{ \boldmath Acknowledgments}
We thank V.L.~Chernyak for many fruitful discussions.
We are grateful for the 
extraordinary contributions of our \pep2\ colleagues in
achieving the excellent luminosity and machine conditions
which made this work possible.
The success of this project also relies critically upon the 
expertise and dedication of the computing organizations that 
support \babar.
The collaborating institutions wish to thank 
SLAC for its support and the kind hospitality extended to them. 
This work is supported by the
US Department of Energy
and National Science Foundation, the
Natural Sciences and Engineering Research Council (Canada),
Institute of High Energy Physics (China), the
Commissariat \`a l'Energie Atomique and
Institut National de Physique Nucl\'eaire et de Physique des Particules
(France), the
Bundesministerium f\"ur Bildung und Forschung and
Deutsche Forschungsgemeinschaft
(Germany), the
Istituto Nazionale di Fisica Nucleare (Italy),
the Foundation for Fundamental Research on Matter (The Netherlands),
the Research Council of Norway, the
Ministry of Science and Technology of the Russian Federation, and the
Particle Physics and Astronomy Research Council (United Kingdom). 
Individuals have received support from 
CONACyT (Mexico),
the A. P. Sloan Foundation, 
the Research Corporation,
and the Alexander von Humboldt Foundation.


\begin{thebibliography}{99}
\bibitem{BM} G.~Bonneau and F.~Martin, Nucl. Phys. B {\bf 27}, 381 (1971).
\bibitem{Coulomb} C.~Tzara, Nucl. Phys. B {\bf 18}, 246 (1970).
\bibitem{kuhn_pp} H.~Czyz {\em et al.}, Eur. Phys. J. C {\bf 35}, 527 (2004).
\bibitem{DM1} DM1 Collaboration, B.~Delcourt {\em et al.}, Phys. Lett. B {\bf 86}, 395 (1979).
\bibitem{DM2} DM2 Collaboration, D.~Bisello {\em et al.}, Nucl. Phys. B {\bf 224}, 379 (1983);
Z. Phys. C {\bf 48}, 23 (1990).
\bibitem{FENICE} FENICE collaboration, A.~Antonelli  {\em et al.}, Nucl. Phys. B {\bf 517}, 3 (1998).
\bibitem{ADONE73} M.~Castellano {\em et al.}, Nouvo Cim. A {\bf 14}, 1 (1973).
\bibitem{BES} BES Collaboration, M.~Ablikim {\em et al.}, 
Phys.\ Lett.\ B {\bf 630}, 14 (2005).
\bibitem{CLEO} CLEO Collaboration, T.~K.~Pedlar {\em et al.}, 
submitted to Phys.~Rev.~Lett., hep-ex/0510005.
\bibitem{LEAR} PS170 Collaboration, G.~Bardin  {\em et al.}, Nucl. Phys.  B {\bf 411}, 3 (1994).
\bibitem{E760} E760 Collaboration, T.~A.~Armstrong {\em et al.}, Phys. Rev. Lett. {\bf 70}, 1212 (1993).
\bibitem{E835} E835 Collaboration, M.~Ambrogiani {\em et al.}, Phys. Rev. D {\bf 60}, 032002
(1999); M.~Andreotti {\em et al.}, Phys. Lett. B {\bf 559}, 20 (2003).
\bibitem{yexp}\babar\ Collaboration, B.~Aubert {\em et al.},
Phys.\ Rev.\ Lett.\  {\bf 95}, 142001 (2005).
\bibitem{ref:babar-nim} \babar\ Collaboration, B.~Aubert {\em et al.},
Nucl. Instr. and Meth. A {\bf 479}, 1 (2002).
\bibitem{EVA} H.~Czyz and J.H.K\"uhn, Eur. Phys. J. C {\bf 18}, 497 (2001).
\bibitem{strfun} M.~Caffo, H.~Czyz, and E.~Remiddi,
Nuo. Cim. {\bf 110A}, 515 (1997);
Phys. Lett. B {\bf 327}, 369 (1994).
\bibitem{PHOTOS}
E.~Barberio and Z.~Was, Comput. Phys. Commun. {\bf 79}, 291 (1994).
\bibitem{JETSET} T.~Sj\"ostrand, Comput. Phys. Commun. {\bf 82}, 74 (1994).
\bibitem{GEANT4} S.~Agostinelli {\em et al.},
Nucl. Instr. and Meth. A {\bf 506}, 250 (2003).
\bibitem{Chernyak} V.L.Chernyak, private communication.
\bibitem{ppres} E687 Collaboration, P.L.~Frabetti  {\em et al.},  
Phys. Lett. B {\bf 578}, 290 (2004).
\bibitem{QCD} V.L.~Chernyak, A.R.~Zhitnitsky, JETP Lett. {\bf 25}, 510 (1977);
G.~Lepage, S.~Brodsky, Phys. Rev. Lett. {\bf 43}, 545, (1979).
\bibitem{ivanch} M.~Benayoun {\em et al.}, Mod. Phys. Lett. A {\bf 14}, 2605 (1999).
\bibitem{psipp1}
DM2 Collaboration, D.~Pallin {\em et al.}, Nucl. Phys. B {\bf 292}, 653 (1987);
DASP Collaboration, R.~Brandelik {\em et al.}, Z. Phys. C {\bf 1}, 233 (1976);
MARKI Collaboration, I.~Peruzzi {\em et al.}, Phys. Rev. D {\bf 17}, 2901 (1978);
MARKII Collaboration, M.W.~Eaton {\em et al.},  Phys. Rev. D {\bf 29}, 804 (1984).
\bibitem{BESpp}BES Collaboration, J.Z.~Bai {\em et al.}, Phys. Lett. B {\bf 591}, 42 (2004).
\bibitem{phokhara} G.~Rodrigo {\em et al.}, 
Eur. Phys. J. C {\bf 24}, 71 (2002).
\bibitem{DE} M.~Davier {\em et al.}, 
Eur.  Phys. J. C {\bf 27}, 497 (2003).
\bibitem{E835psi}E835 Collaboration, M.~Ambrogiani {\em et al.}, 
Phys. Lett. B {\bf 610}, 177 (2005).
\bibitem{pdg} Review of Particle Physics, Phys. Lett. B {\bf 592}, 1, (2004).
\bibitem{E760psi}E760 Collaboration,
T.~A.~Armstrong {\em et al.}, Phys. Rev. D {\bf 47}, 772 (1993).
\bibitem{CLEOpsi}CLEO Collaboration, T.~K.~Pedlar {\em et al.},
Phys. Rev. D {\bf 72}, 051108 (2005).
\bibitem{yhybrid}F.~E.~Close and P.~R.~Page,
Phys.\ Lett.\ B {\bf 628}, 215 (2005).
\bibitem{y4quark}  L.~Maiani, V.~Riquer, F.~Piccinini and A.~D.~Polosa,
Phys.\ Rev.\ D {\bf 72}, 031502 (2005).     
\bibitem{yhybrid0}   S.~L.~Zhu,
Phys.\ Lett.\ B {\bf 625}, 212 (2005).
\bibitem{ymol1} X.~Liu, X.~Q.~Zeng and X.~Q.~Li,
Phys.\ Rev.\ D {\bf 72}, 054023 (2005).
\bibitem{ymol2}  C.~F.~Qiao, hep-ph/0510228.
\bibitem{Neyman} J.~Neyman, Phil. Trans. Royal Soc. London, Ser. A, {\bf 236}, 333 (1937).
\end{thebibliography}
\end{document}